\documentclass{aa}

\usepackage{graphicx}
\usepackage{txfonts}
\usepackage{natbib}        
\bibpunct{(}{)}{;}{a}{}{,} 

\newcommand{\corrections}[1]{#1}

\begin{document}
\title{PIONIER: a 4-telescope visitor instrument at VLTI\thanks{Based on observations collected at the European Southern Observatory, Paranal, Chile (commissioning data and 087.C-0709).}}
\titlerunning{PIONIER: a 4-telescope visitor instrument at VLTI}
\author{
  J.-B.~Le~Bouquin\inst{1}
  \and J.-P.~Berger\inst{2}
  \and B.~Lazareff\inst{1}
  \and G.~Zins\inst{1}
  \and P.~Haguenauer\inst{2}
  \and L.~Jocou\inst{1}
  \and P.~Kern\inst{1}
  \and R.~Millan-Gabet\inst{3}
  \and W.~Traub\inst{4}
  \and O.~Absil\inst{6}\fnmsep\thanks{Postdoctoral Researcher F.R.S.-FNRS (Belgium)}
  \and J.-C. Augereau\inst{1}
  \and M.~Benisty\inst{5}
  \and N.~Blind\inst{1}
  \and X.~Bonfils\inst{1}
  \and \corrections{P.~Bourget\inst{2}}
  \and A.~Delboulbe\inst{1}
  \and P.~Feautrier\inst{1}
  \and M.~Germain\inst{1}
  \and P.~Gitton\inst{2}
  \and D.~Gillier\inst{1}
  \and M.~Kiekebusch\inst{7}
  \and J.~Kluska\inst{1}
  \and J.~Knudstrup\inst{7}
  \and P.~Labeye\inst{8}
  \and J.-L.~Lizon\inst{7}
  \and J.-L.~Monin\inst{1}
  \and Y.~Magnard\inst{1}
  \and F.~Malbet\inst{1}
  \and D.~Maurel\inst{1}
  \and F.~M{\'e}nard\inst{1}
  \and M.~Micallef\inst{1}
  \and L.~Michaud\inst{1}
  \and G.~Montagnier\inst{2}
  \and S.~Morel\inst{2}
  \and T.~Moulin\inst{1}
  \and K.~Perraut\inst{1}
  \and D.~Popovic\inst{7}
  \and P.~Rabou\inst{1}
  \and S.~Rochat\inst{1}
  \and \corrections{C.~Rojas\inst{2}}
  \and F.~Roussel\inst{1}    %
  \and A.~Roux\inst{1}
  \and E.~Stadler\inst{1}
  \and S.~Stefl\inst{2}
  \and E.~Tatulli\inst{1}
  \and N.~Ventura\inst{1}
}

\newcommand\deltaSco{$\delta$~Sco}

\institute{
  UJF-Grenoble 1 / CNRS-INSU, Institut de Plan{\'e}tologie et d'Astrophysique de Grenoble (IPAG) UMR 5274, Grenoble, France
  \and
  European Organisation for Astronomical Research in the Southern Hemisphere (ESO), Casilla 19001, Santiago 19, Chile
  \and
  NASA Exoplanet Science Institute (NExScI), California Institute of Technology, Pasadena, California USA
  \and
  Jet Propulsion Laboratory, California Institute of Technology, Pasadena, California, USA
  \and
  Max Planck Institut f\"ur Astronomie, Konigst\"uhl 17, 69117 Heidelberg, Germany 
  \and
  Institut d'Astrophysique et de G\'eophysique, Universit\'e de Li\`ege, 17 All\'ee du Six Ao\^ut, B-4000 Li\`ege, Belgium
  \and
  European Organisation for Astronomical Research in the Southern Hemisphere (ESO), Karl-Schwarzschild-Str. 2, 85748, Garching bei Munchen, Germany,
  \and
  CEA-LETI, MINATEC Campus, 17 rue des Martyrs, 38054 Grenoble Cedex 9, France
}
\offprints{J.B.~Le~Bouquin\\
  \email{jean-baptiste.lebouquin@obs.ujf-grenoble.fr}}
\date{Received 28/06/2011; Accepted 07/09/2011}
\abstract
{PIONIER stands for Precision Integrated-Optics Near-infrared Imaging ExpeRiment. It combines four $1.8\,$m Auxilliary Telescopes or four $8\,$m Unit Telescopes of the Very Large Telescope Interferometer (ESO, Chile) using an integrated optics combiner. The instrument has been integrated at IPAG starting in December 2009 and commissioned at the Paranal Observatory in October 2010. It provides scientific observations since November 2010.}
{In this paper, we detail the instrumental concept, we describe the standard operational modes and the data reduction strategy. We present the typical performance and discuss how to improve them.}
{This paper is based on laboratory data obtained during the integrations at IPAG, as well as on-sky data gathered during the commissioning at VLTI. We illustrate the imaging capability of PIONIER on the binaries \deltaSco{} and HIP11231.}
{PIONIER provides 6 visibilities and 3 independent closure phases in the H band, either in a broadband mode or with a low spectral dispersion (R=40), using natural light (i.e. unpolarized). The limiting magnitude is $H\mathrm{mag}=7$ in dispersed mode under median atmospheric conditions (seeing$\,<1''$, $\tau_0>3\,$ms) with the $1.8\,$m Auxiliary Telescopes. We demonstrate a precision of $0.5\,$deg on the closure phases. The precision on the calibrated visibilities ranges from 3 to 15\% depending on the atmospheric conditions.}
{PIONIER has been installed and successfully tested as a visitor instrument for the VLTI. It permits high angular resolution imaging studies at an unprecedented level of sensitivity. \corrections{The successful combination of the four $8\,$m Unit Telescopes in March 2011 demonstrates that VLTI is ready for 4-telescope operation.}}

\maketitle

\section{Introduction}

The Very Large Telescope Interferometer \citep[VLTI,][]{Haguenauer:2010} is equipped with four Unit Telescopes of $8\,$m (UTs) and four relocatable Auxiliary Telescopes of $1.8\,$m (ATs). It offers a unique combination of interferometric imaging capability \emph{and} high sensitivity. A nice review of the scientific opportunities is being published by Berger et al. (2011, sub. to A\&A~Rev). It emphasizes the need to recombine a large number of telescopes to provide reliable snapshot imaging capabilities. However, the current instrumentation suite can only handle two or three telescopes simultaneously. The next generation of facility instruments that will combine four telescopes is not expected to be operational prior to 2014.

\begin{figure}
  \centering
  \includegraphics[width=0.49\textwidth]{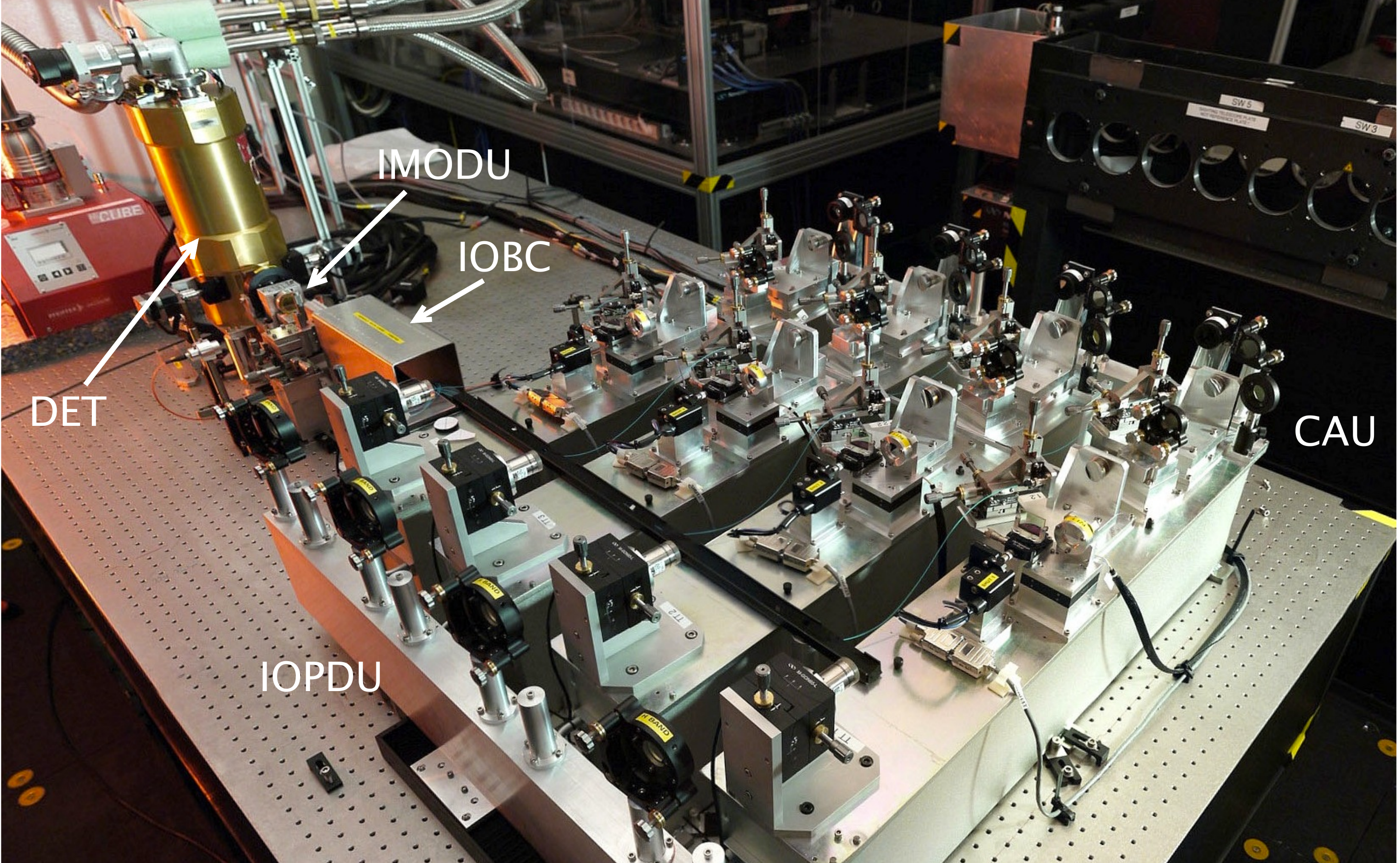}
  \caption{Picture of the PIONIER instrument installed in the VLTI laboratory in October 2010. The different modules refer to the conceptual description of Section~\ref{sec:description} and Fig.~\ref{fig:concept}. The size of the optical table is approximately $1.5\times3\,$m.}
  \label{fig:atvlti}
\end{figure}

Therefore, IPAG\footnote{Institut de Plan{\'e}tologie et d'Astrophysique de Grenoble} and its partners have proposed to use a \emph{visitor instrument}\footnote{http://www.eso.org/sci/facilities/paranal/instruments/vlti-visitor} that provides VLTI, since 2010, with a new observational capability combining imaging, sensitivity and precision. The principle of this Precision Integrated-Optics Near-infrared Imaging ExpeRiment (PIONIER) was approved by the ESO Science and Technical Commitee in spring 2009. The instrument has been integrated at IPAG starting in December 2009 and commissioned at the Paranal Observatory in October 2010. Only a few nights of commissioning were necessary before the instrument started routinely delivering scientific data with the ATs. \corrections{On 17 March 2011 light collected by all four of the $8\,$m UTs was successfully combined for the first time.} A picture of PIONIER installed at the VLTI laboratory is shown in Fig.~\ref{fig:atvlti}. 

PIONIER relies on the integrated optics (IO) technology to combine four beams \citep{malbet:1999jul,berger:2001sep,Benisty:2009may}. In one observation, it provides the measurements of six visibilities and three independent closure phases with low spectral resolution ($R\approx40$) in the H band. In many respect, it is complementary to the current AMBER instrument that provides three visibilities and one closure phase with various spectral resolutions ($R\approx30,1\,500,12\,000$) in the H and K bands \citep{petrov:2007mar}. PIONIER paves the road for the forthcoming instruments GRAVITY and MATISSE \citep{Gillessen:2010,Lopez:2008}, by enabling the 4-telescope operation of VLTI. 

The paper is organized as follow: section~\ref{sec:science} presents the main astrophysical programs of PIONIER and the corresponding high-level specifications. Section~\ref{sec:description} describes the instrument and its different subsystems. Section~\ref{sec:operation} presents the PIONIER operations, including regular day-time maintenance, flux injection, scientific fringe tracking and regular calibrations. Section~\ref{sec:drs} presents the data reduction strategy, focusing on the aspects that are specific to the instrument, or that have not been published elsewhere, and the current performance. Finally, Section~\ref{sec:illustration} illustrates the imaging capability of PIONIER on the Be star 
\deltaSco{} and the spectroscopic binary HIP11231. The paper ends with brief conclusions and remarks about possible evolutions of the instrument.

\section{Science drivers and specifications}
\label{sec:science}
PIONIER is designed for aperture synthesis imaging with a specific emphasis on fast fringe recording to allow precision closure phases and visibilities to be measured. Among all the possible science cases, we defined three key astrophysical programs.

\subsection{Young Stellar Objects}
Young stars are surrounded by an accretion disk made of gas and dust, thought to be the birthplace of planets. Studying the innermost astronomical unit is of  major importance to understand the physical conditions for star and planet formation.

Herbig Ae/Be stars are young stars of intermediate-mass, whose disks have been intensively studied with near-infrared interferometers in the last 13 years.  Measurements enabled the determination of near-infrared sizes, found not to agree with classical accretion disk model predictions (see \citet{Millan-Gabet:2007} for a review). Consequently the view on the inner structure of accretion disks has been substantially revised. The concept of ``puffed-up'' inner rim has been introduced \citep{Dullemond:2001,Isella:2005}. It marks the boundary in the disk where dust can no longer exist due to sublimation and forms a barrier to the photons potentially causing an alteration of the vertical structure of the disk. Recently, strong evidence was found that the matter flowing through this limit to the star is a major contributor to the excess emission \citep{Tannirkulam:2008,Benisty:2010}. Revealing the nature of the matter inside the dust sublimation limit and the exact structure of the inner disk at this transition is now investigated with care. We have identified a small sample of HAe/Be stars whose environments have been clearly resolved by the VLTI and should therefore be ideal candidates for imaging with PIONIER in order to reveal the morphology of the inner astronomical unit near-infrared emission. \emph {The simultaneous combination of four telescopes is required to obtain reliable snapshot image reconstructions of these complex and time-variable objects \citep{Benisty:2011}.}

In addition, PIONIER will probe the structure of disks around the T~Tauri stars, which are the low-mass analogs of Herbig stars. In these fainter objects, the inner edge of the dusty disk is closer to the star, and is not expected to be fully resolved, preventing us from direct imaging. However, the visibilities will give access to the inner disk location \citep{Akeson:2005,Eisner:2009} and the closure phases will probe the morphology and surface brightness profile. In the best cases, it will be possible to constrain, through proper radiative transfer modeling, the inner rim position, ellipticity and thickness. All of these are of major importance to understand the conditions for star and planet formation. \emph{As T~Tauri stars are the faintest objects of our key programs, they set the need for limiting magnitude of $H\mathrm{mag}>7$.}

\subsection{Debris disks around main sequence stars}
\label{sec:debrisdisks}
The inner solar system contains a cloud of small dust grains created when small bodies --asteroids and comets-- collide and outgas. This dust cloud, called the zodiacal disk, has long been suspected to have extrasolar analogs around main sequence stars. The very inner part of these exozodiacal disks have remained elusive until recently, when near-infrared interferometric observations revealed a resolved excess flux of about 1\% around several nearby stars \citep{absil:2006jun,Absil:2008sep,Absil:2009,di-Folco:2007,Akeson:2009}. These excesses have been interpreted as being due to hot, possibly transient exozodiacal dust within $1\,$AU from these stars.

Our main goal is to survey main sequence stars to study the hot dust content during the late phases of terrestrial planet formation, and its connection with outer resevoirs of cold dust. We expect that about 100 main sequence stars of various ages between $10\,$Myr and a few Gyr will be observed during the PIONIER lifetime, providing us with a solid statistical sample to derive the time dependence of the bright exozodi phenomenon. In addition, for the brightest cases, PIONIER could be used to characterize the grain properties thanks to multi-colour information, and to study the inner disk morphology. \emph{The challenge of this program is to obtain an accuracy of 1\% or less on the calibrated visibility.}

\subsection{Binaries, faint companions and Hot Jupiters}
\corrections{Long baseline interferometric observations naturally complement the radial velocity and adaptive optics surveys to discover and characterize binary systems. For instance, radial velocity measurements on massive stars are challenged by the lack of spectral lines and their intrinsic broadening. On active stars, the intrinsic radial velocity jitter may easily hide the signal of a faint companion. Moreover, compared with adaptive optics, the access to smaller separations and therefore shorter periodes increases the possibility to determine the dynamical masses of the components.

Our first attempts with PIONIER (Absil~et~al. 2011, submitted) demonstrate that the angular separation range from $5$ to $100\,$mas can be covered within one hour of observation, thanks to the simultaneous measurement of 4 closure phases at a low spectral resolution. Accordingly, we have initiated several surveys from massive to low mass stars, including young stars and stars in nearby moving groups. \emph{In order to reach a sufficient number of targets, these programs typically require a limiting magnitude of $H\mathrm{mag}=6$ and a precision of $0.5\deg$ on the closure phases (1:200 dynamic range).}}

Deep integration on few selected targets will permit to assess what is the best possible dynamic range achievable, until possibly reaching the planetary regime (1:2000). If successful PIONIER will have the means to separate the flux contribution of the planet from that of its parent star and potentially obtain low spectral resolution information. \emph{In order to achieve such a result, systematic biases on the closure phase estimate have to be hunted down and understood at the unprecedented level of $0.05\,$deg \citep{Zhao:2010,Zhao:2011}.}

\subsection{High level specifications}
\begin{table}
  \centering
  \caption{Summary of the scientific requirements compared to the already demonstrated performance. Error is defined as the accuracy for one calibrated data point.}
  \begin{tabular}[c]{lccccccc} \hline \hline \vspace{-0.3cm}\\
    Topic & Sp. Band & Mag. & $V^2$ error & CP error\\\hline
    Herbig AeBe disks & H,K & $>5$ & 5\% & 5deg\\
    T~Tauri disks & H,K & $>7$ & 5\% & 2deg\\
    Debris disks & H,K & - & 1\% & 1deg\\
    Faint companions & H,K &$>6$&- & 0.5deg\\
    Hot Jupiters & H,K &-&- & 0.05deg\\\hline
    Demonstrated & H & 7.5 (AT) & 15 -- 3\% & 0.5deg\\\hline
  \end{tabular}
  \flushleft
  \label{tab:obs_log}
\end{table}
The specifications of PIONIER have been defined as a trade-off between the requirements for the key astrophysical programs and the technical contraints due to the project small size and fast timeline. A summary is presented in Table~\ref{tab:obs_log}. We used the experience gained with IONIC-3 \citep{berger:2003feb}, VINCI \citep{le-bouquin:2004sep,le-bouquin:2006may}, and AMBER to make the following strategic choices:
\begin{itemize}
\item the combination of 4 telescopes (6 baselines) simultaneously,
\item the use of an integrated optics beam combiner,
\item operation in the H- and K-band (not simultaneously),
\item a small spectral dispersion across few channels,
\item a fast readout of the camera.
\end{itemize}

As built, PIONIER matches all these requirements except for the K band extension which is contemplated in the future. The instrument has been integrated and commissioned with the H-band integrated beam combiner that was already available at the IPAG laboratory \citep{Benisty:2009may}.

\section{Instrument description}
\label{sec:description}

\begin{figure*}
  \centering
  \includegraphics[width=\textwidth]{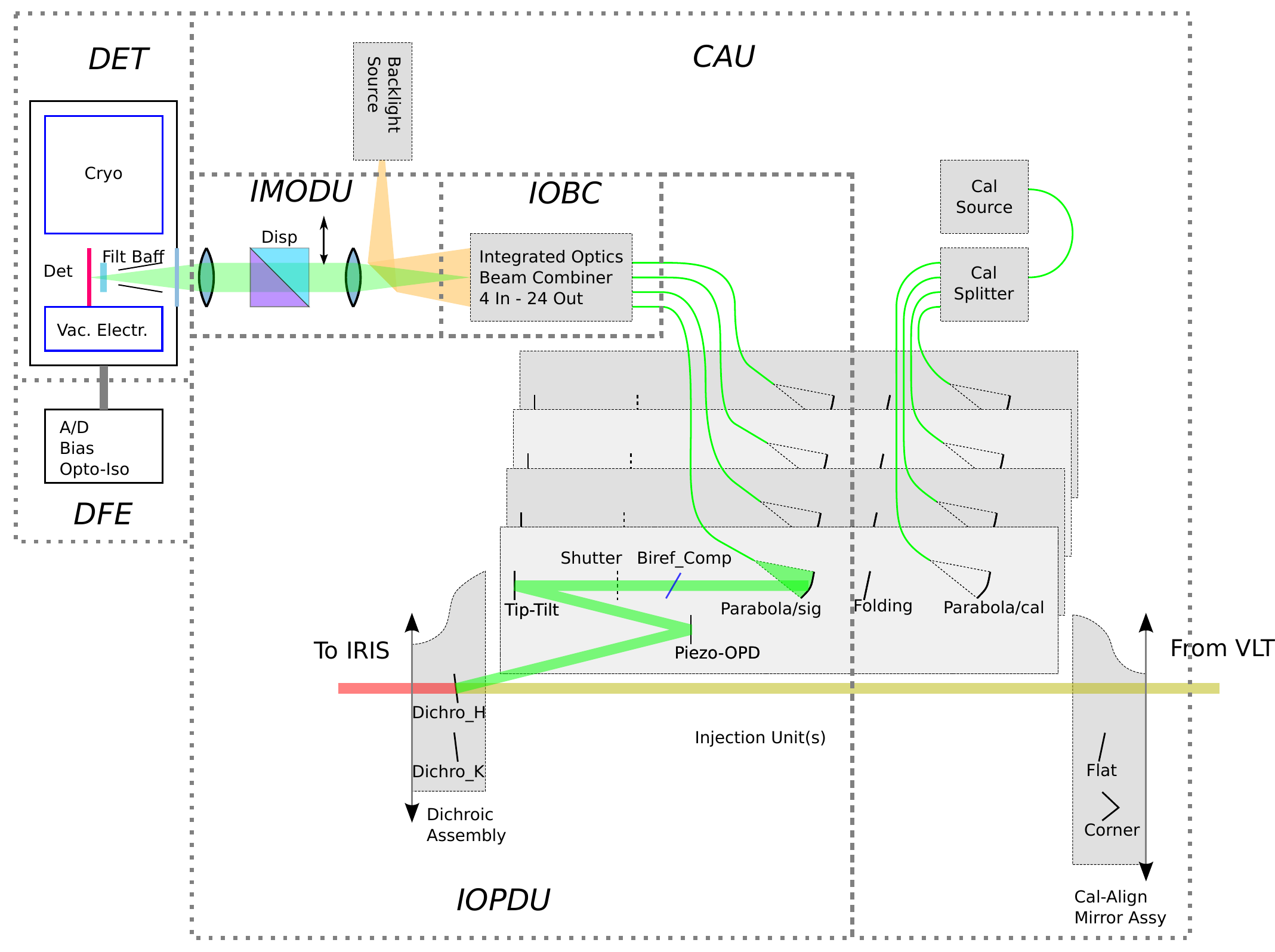}
  \caption{PIONIER conceptual scheme. The Injection and Optical Delay Unit (IOPDU) is responsible for the injection of the four beams (either from CAU or from VLTI) in the single mode fibers. It also implements the OPD modulation, the tip-tilt correction and the polarization control. The Integrated Optics Beam Combiner (IOBC) recombines the four beams into 24 interferometric outputs in a pairwise scheme. These outputs are then imaged, and eventually spectrally dispersed, by the Imaging Optics and Dispersion Unit (IMODU) on the detector (DET). The aim of the Calibration and Alignment Unit (CAU) is to provide four interferometric beams that match the size and angle of the VLTI beams.}
  \label{fig:concept}
\end{figure*}

Figure~\ref{fig:concept} summarizes the key elements of the PIONIER instrument. Functionally, it consists of the following subsystems.

\subsection{Injection and Optical Delay Unit (IOPDU)}
\label{sec:iopdu}
The Injection and Optical Delay Unit (IOPDU) injects the free-space beams from the VLTI into optical fibers.  It includes a tip-tilt correction, an OPD modulation and a polarization control. The IOPDU is made of four strictly identical arms, one per VLTI beam. 
\begin{enumerate}
\item A dichroic mirror extracts the required band from the VLTI beams, while other wavelengths are used to feed the infrared VLTI guiding camera IRIS. The translation stage that supports these optics has two observing positions (the H-band and K-band dichroic) and a free position to let the beam pass unaffected when PIONIER is not in use.
\item A modulation of the optical path length is introduced by a mirror mount in a Physik Instrument (PI) piezo translation stage. It provides an OPD range of $\pm 400\,\mu$m.
\item The tip-tilt mirror, mounted on PI piezo devices, allows the beam angle to be corrected up to a frequency of $100\,$Hz with a range of $\pm200"$ (laboratory angle, corresponding to $\pm13$ times the size of the magnified PSF).
\item \corrections{A Lithium-Niobate plate of 2\,mm thickness is used to compensate for the polarization phase-shift arising in the single-mode fiber of the IOBC (see Sec.~\ref{sec:iobc}). The amount of phase-shift between the vertical and horizontal axes is accurately adjusted by tilting differentially the plates in the individual beams.}
\item The off-axis parabola focuses the light into the single mode, polarization maintaining fiber.
\end{enumerate}

PIONIER's fast tip-tilt correction comes in addition to the one provided by the telescopes. It mainly compensates for the additional contributions coming from the tunnel turbulence. The IRIS guiding camera provides the beam angle of arrival measurements at a rate from $100\,$Hz to $1\,$Hz depending on the target brightness \citep{Gitton:2004}.

\subsection{The Integrated Optics Beam Combiner (IOBC)}
\label{sec:iobc}
The Integrated Optics Beam Combiner (IOBC) takes as input the signals from the IOPDU, and delivers 24 interferometric outputs.

The IOBC is fed by polarization maintaining single-mode H-band fibers, whose lengths have been equalized with an accuracy of $20\,\mu$m. Consequently, these fibers introduce a negligible differential chromatic dispersion. The polarization phase-shift between the vertical and horizontal axes (i.e. the neutral axes of the IO chip, aligned with those of the fibers) is of the order of 1 fringe.  \corrections{This polarization phase-shift can be compensated by adjusting the inclination of the Lithium-Niobate plate in the IOPDU (see Sec.~\ref{sec:iopdu}).}

The design of the IOBC is described in detail in \citet{Benisty:2009may}. The four incoming beams are split in three and distributed in the circuit in a pairwise combination. A so-called ``static-ABCD'' combining cell is implemented for each baseline. It generates simultaneously four phase states (almost in quadrature). Consequently, 24 outputs have to be read. The low-chromaticity phase shift is obtained through the use of specific waveguides with carefully controlled refraction index. This combiner can be used both in fringe-scanning mode (VINCI-like) or ABCD-like mode. There is currently one combiner available in the H band and one that is being developed for the K band \citep{Jocou:2010}.

\subsection{Imaging Optics and Dispersion Unit (IMODU)}
The Imaging Optics and Dispersion Unit (IMODU) images the 24 outputs of the IOBC onto the camera's focal plane, with or without spectral dispersion in the perpendicular direction (1, 3 or 7 spectral channels across the H band). This unit is the one that was used with the IONIC-3 instrument at IOTA \citep{berger:2003feb}. Its image quality permits to focus about 80\% of the flux into a single pixel of the PICNIC camera (pixel size $40\,\mu$m). A Wollaston prism can be inserted to acquire separately the fringes for the two linear polarizations (vertical and horizontal axes). \corrections{This device is only used to internally adjust the compensation of the polarization phase-shift (see Sec.~\ref{sec:iobc}). Once this adjustment is done, the Wollaston prism is removed and the fringes are acquired in natural (unpolarized) light without loss of contrast. Consequently the instrument provides no polarization information in targets.}

In practice, these prisms are inserted and removed manually from the optical beam. The positioning of the Wollaston prism is repeatable at $\pm0.1$~pixel. The positioning of the dispersive prism is repeatable at $\pm0.1$~pixel. Altogether, these measurements are compatible with a stability of $\pm1\%$ for the effective wavelength and for the flux splitting ratio between the outputs.

\subsection{The detector (DET/DFE)}
The detector (DET) comprises the focal plane array, the cryostat, the filter wheels, and the internal electronics. More electronic functions, including the digitizing of the video signal, are provided by the Detector Frontend Electronics (DFE). The detector is the PICNIC camera previously used at PTI and IOTA \citep{traub:2004oct,Pedretti:2004fk}. Its electronics has been partially upgraded to improve the frame rate. The array is clocked at $5\,$MHz and performs analog-to-digital conversions every $0.4\,\mu$s. The main limitation for speed is the time for the video signal to stabilize once a pixel has been addressed ($\approx 3\,\mu$s). The resulting time for the elementary operations reset and read are presented in Table~\ref{tab:picnictime}. Consequently, the detector achieves a frame-rate of $0.85\,$kHz when reading 168 pixels (24 outputs times 7 spectral channels) in a non-destructive way.

We characterized the readout noise following the same procedure as described by \citet{Pedretti:2004fk}. We found a noise of $\approx 13e^-$ per read while Pedretti et al. reported $\approx 9e^-$ per read on this camera. This noise is reduced to $\approx 8e^-$ by averaging 8 successive analog-to-digital conversions of the same pixel. It is still possible to reduce the noise by averaging several \emph{frames} in a non-destructive way, although it has a strong impact on the minimum integration time possible.

\begin{table}
  \centering
  \caption{\label{tab:picnictime} Typical time to execute elementary operations with our fastest possible clocking of the PICNIC camera. These numbers are for a single analog-to-digital conversion per pixel.}
  \begin{tabular}[c]{lcccc} \hline \hline \vspace{-0.3cm}\\
    Detector elementary & Required time \\
    operation & ($\mu$s) \\\hline
    Reset full quadrant &  205 \\
    Read 24 pixels $\times$ 1 line  & 190 \\
    Read 24 pixels $\times$ 7 lines & 1150 \\\hline
  \end{tabular}\\
\flushleft
{\footnotesize }
\end{table}

\subsection{The Calibration and Alignment Unit (CAU)}
 The Calibration and Alignment Unit (CAU) serves dual functions: (1) to inject, as a substitute for the astronomical beams, mutually coherent beams derived from a common Tungsten source, for laboratory verification and health check purposes; (2) to reflect, via a corner cube mirror, a reverse-propagating beam from the output of the IOBC towards the IRIS guiding camera, for the purpose of verifying the alignment of the PIONIER instrument with the IRIS reference positions.

\subsection{Control System and electronics (CS)}
Control System (CS) includes hardware control of the instrument units, detector readout, quicklook and interaction with VLTI. The hardware has been designed to follow the standard VLTI architecture including an instrument workstation (WS), an Instrument Control System (ICS), a Local Control Unit (LCU) and a Detector Control System (DCS) running on the detector WS. The instrument WS is a standard Linux PC platform. The detector WS is an industrial Linux PC platform equipped with a PCI-7300A board used to generate clock signals and to acquire detector data. Both stations are running under the VLT Software.

Instead of the traditional LCUs, the instrument is controlled by an Embedded PC and several EtherCAT modules like remote IO and stepper motion controllers from the company Beckhoff. Through a fruitful collaboration with ESO, PIONIER has been used as a pilot project for testing these commercial components and developing the associated VLT Software extensions. This success not only tries to simplify the coding of ICS but it also attempts to introduce technologies envisaged for E-ELT instruments \citep{Kiekebusch:2010}.

PIONIER electronics is housed in a cooled cabinet close to the instrument optical table. It includes the detector front-end electronics, the DC detector power supplies together with the detector WS, tip-tilt and scanning piezo controllers, EtherCAT, calibration lamps and control electronics for electro-mechanical devices.

\subsection{Interfaces}
Dialog between PIONIER and VLTI (e.g. sending new coordinates) is achieved through the VLTI Interferometer Supervisor Software. The detector WS is also listening to the VLTI Reflective Memory Network (RMN) to get real-time centroid information from the IRIS guiding camera. 

The raw output data follow the standard defined by ESO for its interferometric instruments \citep{ballester:2002jun}, including the general FITS header. The only difference is the addition of a new dimension (depth) in the column DATAj of the binary table IMAGING\_DATA. The third dimension of DATAj represents the consecutive analog-to-digital conversions of each pixel. Consequently, a typical night provides several gigabytes of data.

The Observation Software running on the instrument workstation coordinates the execution of an exposure for a given observing mode. From the high level point of view, PIONIER is operated via the Broker of Observing Blocks that executes Observing Blocks (OBs) fetched from the standard \texttt{p2pp} ESO software. OBs can be conveniently generated by the \texttt{aspro2}\footnote{http://www.jmmc.fr/aspro\_page.htm} preparation software from the Jean-Marie Mariotti Center (JMMC).

\section{Operating PIONIER}
\label{sec:operation}

This section describes the PIONIER operations, including regular day-time optical alignment, flux injection procedure, scientific fringe recording and regular calibrations.

\subsection{Optical alignment}
As a visitor instrument, PIONIER is maintained and operated by the PIONIER team. It is only operated in visitor-mode and the observer is responsible for checking the health of the instrument at the beginning and during the observing run. The day-to-day stability of the beam angle inside PIONIER is $\approx 4''$, that is a quarter of the magnified PSF size. This is well within the range of the tip-tilt actuators ($\pm200''$). The day-to-day position stability of the fringes acquired with the internal calibration is better than $15\,\mu$m in OPD. The same typical variations are observed with the VLTI calibration source MARCEL. Again, this is well within the range of the modulation mirror ($\pm500\,\mu$m) and does not impact the operations.

However, the two following adjustments should be carefully checked at the beginning of each observing run:\begin{itemize}
\item the vertical/horizontal positioning of the IOBC, in order to optimize the coupling of the output spots into the detector pixels ; 
\item the tilt angle of the Lithium-Niobate plates, as the polarisation phase-shifts of the fibers drift by typically $\lambda/8$ over a month timescale.
\end{itemize}
These adjustments remain valid for the few following days of a typical run.

\begin{figure*}
  \centering
  \includegraphics[width=0.9\textwidth]{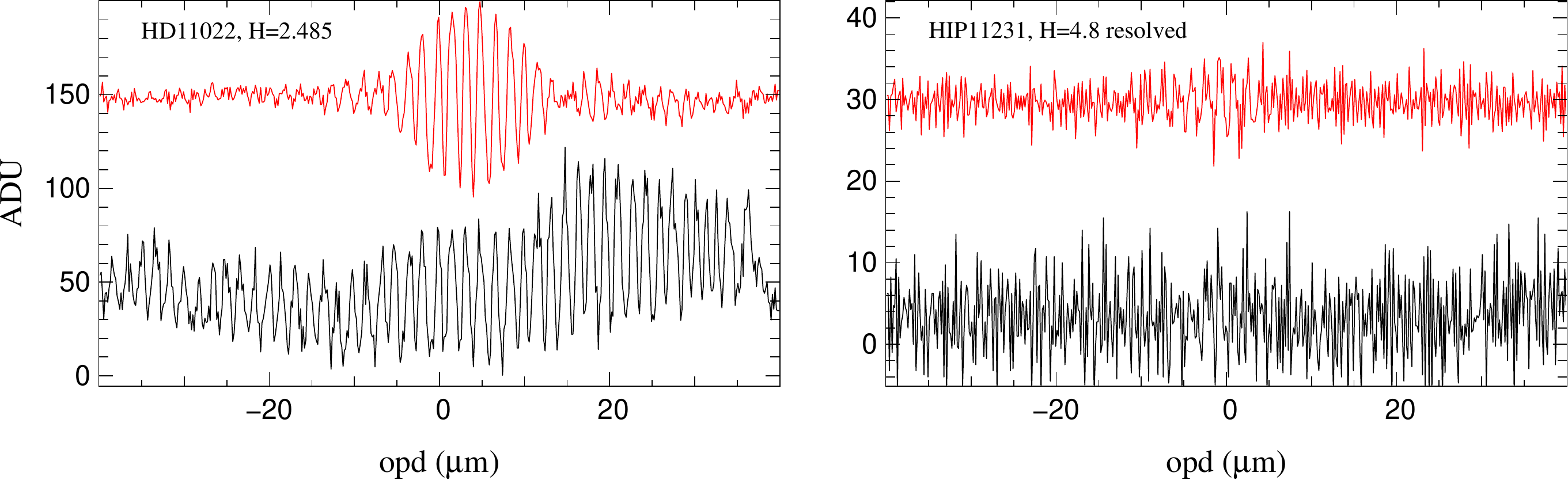}
  \caption{Scanned interferograms obtained with PIONIER on the high SNR regime (left) and the low SNR regime (right) and using the spectral dispersion of 7 channels across the H-band. An example with the baseline with the smallest frequency is represented here. The bottom, black signal represents one output of the combiner, for the central spectral channel (i.e. one pixel of the detector), versus the optical path difference in the scan. The large temporal coherence and the photometric fluctuations are clearly visible. The top, red signal represents the subtraction of two conjugate outputs, after averaging the 7 spectral channels. The SNR is enhanced and the photometric fluctuations are almost perfectly removed.}
  \label{fig:interferograms}
\end{figure*}

\subsection{Flux injection}
\label{sec:on-sky}

The stellar images are stabilized in tip-tilt by the IRIS guiding camera located in the VLTI laboratory \citep{Gitton:2004}, only a few meters away from PIONIER. \corrections{The image shift caused by the difference of atmospheric refraction index between the H-band (PIONIER fibers) and the K-band (IRIS guiding camera) is handled by the Interferometer Supervisor Software. The software automatically offsets the IRIS guiding point so that the stellar image in the H-band is kept fixed whatever the airmass and pointing direction.}

The amount of flux injected in the fibers is optimized by a grid search with the internal tip-tilt of PIONIER. We use a square grid of 10x10 steps of $2.5''$ each. The resulting image is fitted by a 2D gaussian. The procedure is applied sequentially on the four beams, during the evening twilight, using the beacon sources located on the VLTI telescopes. The resulting accuracy is better than $1/50^{\mathrm{th}}$ of the PSF size. The precision is dramatically reduced by speckle noise when the procedure has to be performed on stellar light.

\subsection{The interferometric signal}
The fringes are temporally scanned across the coherence length by means of the long-range OPD modulation devices. Figure \ref{fig:interferograms} shows examples of raw interferograms obtained in the high SNR and low SNR regimes.
The four beams are modulated at non redundant velocities $(-3v,-v,+v,+3v)$ in order to generate the six fringe signals. The speed $v$ is defined by the ratio between the length of the smallest scan (set to $80\,\mu$m) and the duration of a scan. For faint stars, its optimal value is a compromise between sensitivity and the necessity to freeze the atmospheric effects for the fringes at the lowest frequency $f_{min}=2v/\lambda$. For bright stars, $v$ is imposed by the detector frame rate and the need to correctly sample the fringe at the highest frequency $f_{max}=6v/\lambda$.

The most used scan setups are summarized in Table~\ref{tab:fringefreq}. In mode ``fowler'', the detector is reset once and is left integrating non destructively during the entire scan. This mode achieves the best frame rate and the lowest readout noise (when oversampling the fringes). In mode ``double'', the detector performs a reset-read-read sequence for each step in scan. This mode is used on bright stars that would saturate the detector in ``fowler'' mode. Its main drawback is to slow down dramatically the frame rate. \corrections{Consequently, it is not possible to use the mode ``double'' with the spectral dispersion over 7 channels as the resulting frame rate would be too slow compared with the atmospheric coherence time.}

\begin{table}
  \centering
  \caption{\label{tab:fringefreq} Fringe frequencies versus the number of spectral channels to be read and the number of steps in scan. The scan length corresponding to $f_{min}$ is set to $80\,\mu$m. The modes are sorted from less sensitive (top) to most sensitive (bottom).}
  \begin{tabular}[c]{ccccc} \hline \hline \vspace{-0.3cm}\\
    Detector & \# of spectral & steps in& $f_{min}$ &steps per fringe\\
    mode     &  channels      & scan    & (Hz) & at $f_{max}$ \\\hline
    Double   & 1 & 512  & 160 & 3.6 \\
    Fowler   & 7 & 512  & 80  & 3.6 \\ 
    Fowler   & 7 & 1024 & 40  & 7.2 \\
    Fowler   & 1 & 1024 & 240 & 7.2 \\
    Fowler   & 1 & 2048 & 120 & 14.4 \\\hline
  \end{tabular}\\
\flushleft
{\footnotesize }
\end{table}

\subsection{Quick-look and fringe tracking}
\label{sec:tracking}
During the observation, each scan is processed by a quick-look algorithm that also implements a slow group-tracking technique:\begin{enumerate}
\item The spectral channels are added together in a pseudo broad-band signal. The four outputs of each baseline are summed together taking into account their relative phase shift. The result is a single interferometric signal per baseline represented in red in Fig.~\ref{fig:interferograms}.
\item At the end of the scan, the signal is processed by the algorithm presented by \citet{pedretti:2005sep}. It provides a SNR and a group-delay position for each baseline.
\item Then these six group-delay measurements are converted into offsets for the four input beams. The inversion algorithm only takes into account the baselines with an SNR larger than 3. The redundancy is used to recover the offsets of a maximal number of beams. The results are an optimal estimate of the offsets to be applied to the beams and a detection flag per beam (0/1).
\item For all beams with a positive detection flag, the zero positions of the scanning piezo-electric devices is updated with the new offset value, before the next scan starts. 
\end{enumerate}
The typical repetition rate of the loop ($1\,$Hz) is enough to keep the group-delay residuals within $\pm 20\,\mu$m under median atmospheric conditions. The redundancy of the 6 baselines to recover 3 optical path differences makes the tracking robust against temporary flux losses or strongly resolved baselines.

\subsection{Observations and calibration}
\label{sec:calibration}
\corrections{Each pointing typically produces 5 files of 100 fringe scans.} Then a dark exposure is recorded in a file of 100 scans with the same detector setup but with all shutters closed. The flux splitting ratios of each input are finally calibrated by files of 50 scans with one beam illuminated at a time.

The sequence takes between $5\,$min and $10\,$min depending on the modulation speed. An additional 3 min are needed to preset the VLTI telescopes and delay lines and acquire the star. A single calibrated point (science + calibrator) takes between $15$ to $30\,$min. It is possible to skip the optimisation of the telescope guiding system if the V~magnitudes of the science and calibration stars are within $0.5\,$mag. It significantly speeds up the target re-acquisition ($1\,$min instead of $3\,$min).

The spectral calibration is performed at the end of the night. It consists of long scans of $200\,\mu$m with high SNR recorded in the internal CAU. The effective wavelength is estimated by Fourier Transform Spectrometry. We checked the linearity of the OPD modulation by observing a $1.55\,\mu$m infrared laser. The day-to-day repeatability of the effective wavelength is $\pm1\%$, which is consistent with the optical stability presented in the previous section. Taking into account the systematic uncertainties, the final accuracy on the effective wavelength is $\pm2\%$. \corrections{It matches the requirement of the key programs presented in Sec.~\ref{sec:science}. In particular, it meets the required 1\% accuracy on the calibrated visibility associated to the most demanding program, for which the visibility models are not very sensitive to the wavelength calibration (Sec~\ref{sec:debrisdisks}: faint debris disks around marginally resolved photospheres).}

\section{Data reduction and performances}
\label{sec:drs}
In this section, we focus on the steps of the data reduction that are specific to the PIONIER instrument, or that have not already been published.

The data reduction software of PIONIER converts the raw FITS file produced by the instrument into calibrated visibilities and closure phase measurements written in the standard OIFITS format \citep{Pauls:2005fk}. These output files are science-ready and can be directly handled by software such as LITpro \citep[model fitting,][]{Tallon-Bosc:2008} or MIRA \citep[image reconstruction,][]{Thiebaut:2008}. The PIONIER data reduction software is written as a \texttt{yorick}\footnote{{http://yorick.sourceforge.net}} package called \texttt{pndrs} and publicly available\footnote{{http://apps.jmmc.fr/\~{}swmgr/pndrs}}.

Our strategy is directly inspired from the FLUOR, VINCI and IONIC-3 experiments. In these instruments, the scanning method associated with an estimate of the visibility in the Fourier space provided sensitive and accurate measurements, without the need for tricky internal calibrations. Basically, the intensity measured at the output $A$ (respectively B, C and D) of the baseline $ij$ of the combiner is written as:
\begin{equation}
i_A^{ij} = \kappa^{ij}_A\,P^i +  \kappa^{ji}_A\,P^j + \sqrt{\kappa^{ij}_A \kappa^{ij}_A P^i P^j} \mu_A^{ij}\,V^{ij}\,\cos(\frac{2\pi\delta^{ij}}{\lambda}+ \phi^{ij}_A + \Phi^{ij})
\end{equation}
where $ij$ are the baselines (12, 13, 14, 23, 24, 34), $P^i$ are the photometric flux from input beams, $\mu_A^{ij}$ and $\phi_A^{ij}$ are the instrumental visibilities and phases of the beam combiner, $\kappa{}$ are the internal flux splitting ratio, and $\delta^{ij}$ is the OPD modulation across the scan plus the atmospheric piston. The complex visibility of the observed target is $V^{ij}\,e^{\mathrm{i}\,\Phi^{ij}}$.

\subsection{Cosmetic}
The basic operations that are related to the PICNIC detector are performed first. The consecutive non-destructive reads are subtracted accordingly to the detector mode (``double'' or ``fowler'') to provide the intensity. The dark level is removed. In the ``fowler'' mode of the detector, a dark level is estimated independently for each step in scan.

The flux splitting ratios are estimated from the dedicated calibration files. They are computed independently for each wavelength. We use these flux splitting ratios to compute what we call the flat-fielded kappa matrix:
\begin{equation}
\tilde{\kappa}_A^{ij} = {\kappa^{ij}_A}/{(\sqrt{\kappa^{ij}_A\kappa^{ji}_A}\mu_A^{ij})}
\end{equation}
and the flat-fielded signal:
\begin{equation}
\tilde{i}_A^{ij} = {i^{ij}_A}/{(\sqrt{\kappa^{ij}_A\kappa^{ji}_A}\mu_A^{ij})}
\end{equation}
We use the instrumental visibilities $\mu_A^{ij}$ coming from laboratory measurements of the beam combiner, actually close to one.

\subsection{Disentangling the photometric and interferometric signals}
The pairwise design allows the instantaneous photometric signal to be extracted by a global fit to the coherent data. The main idea is to add together the outputs $A,B,C$ and $D$ from a given baseline, after flat-fielding:
\begin{equation}
c^{ij} = \tilde{i}_A^{ij} + \tilde{i}_B^{ij} + \tilde{i}_C^{ij} + \tilde{i}_D^{ij}
\end{equation}
The opposite phase relation between $A$ and $C$ (respectively $B$ and $D$) due to energy conservation removes the interferometric part of the signal:
\begin{equation}
c^{ij} = \tilde{\kappa}_{c}^{ij}\,P^{i}\;+\;\tilde{\kappa}_{c}^{ji}\,P^{j} \label{eq:photmatrix}
\end{equation}
with $\tilde{\kappa}_{c}^{ij} = \tilde{\kappa}_A^{ij} + \tilde{\kappa}_B^{ij} + \tilde{\kappa}_C^{ij} + \tilde{\kappa}_D^{ij}$.~Equation~\ref{eq:photmatrix} defines a linear system between the six $c^{ij}$ signals (one per baseline) and the four $P^i$ photometric signals (one per beam). This system can be inverted and is well constrained thanks to the pairwise combination. We obtain an estimate of the $P^i$ for each step in scan. \corrections{This method to recover the photometric signals was used in the PIONIER precursor instrument IONIC-3 \citep{monnier:2004feb}. Although it is not widely known, this method is a fundamental advantage of multi-telescope pairwise combination \citep[see \$\,4.1.2 in][]{Blind:2011}.}

We build a single interferometric signal per baseline combining the information of all outputs. We first inject the $P^i$ estimates into the flat-fielded kappa matrix in order to remove the continuum part of the interferometric signals. Then we co-add the $A$,$B$,$C$ and $D$ outputs taking into account their relative phase shift (estimated from laboratory measurements):
\begin{eqnarray}
m^{ij} &=& \sum_{o=ABCD}(\tilde{i}_{o}^{ij} - \tilde{\kappa}_o^{ij}P^i - \tilde{\kappa}_o^{ji}P^j)\;e^{-\mathrm{i}\,\Phi_o^{ij}}
\end{eqnarray}

\subsection{Closure phase estimate}
\begin{figure}
  \centering
  \includegraphics[scale=0.345]{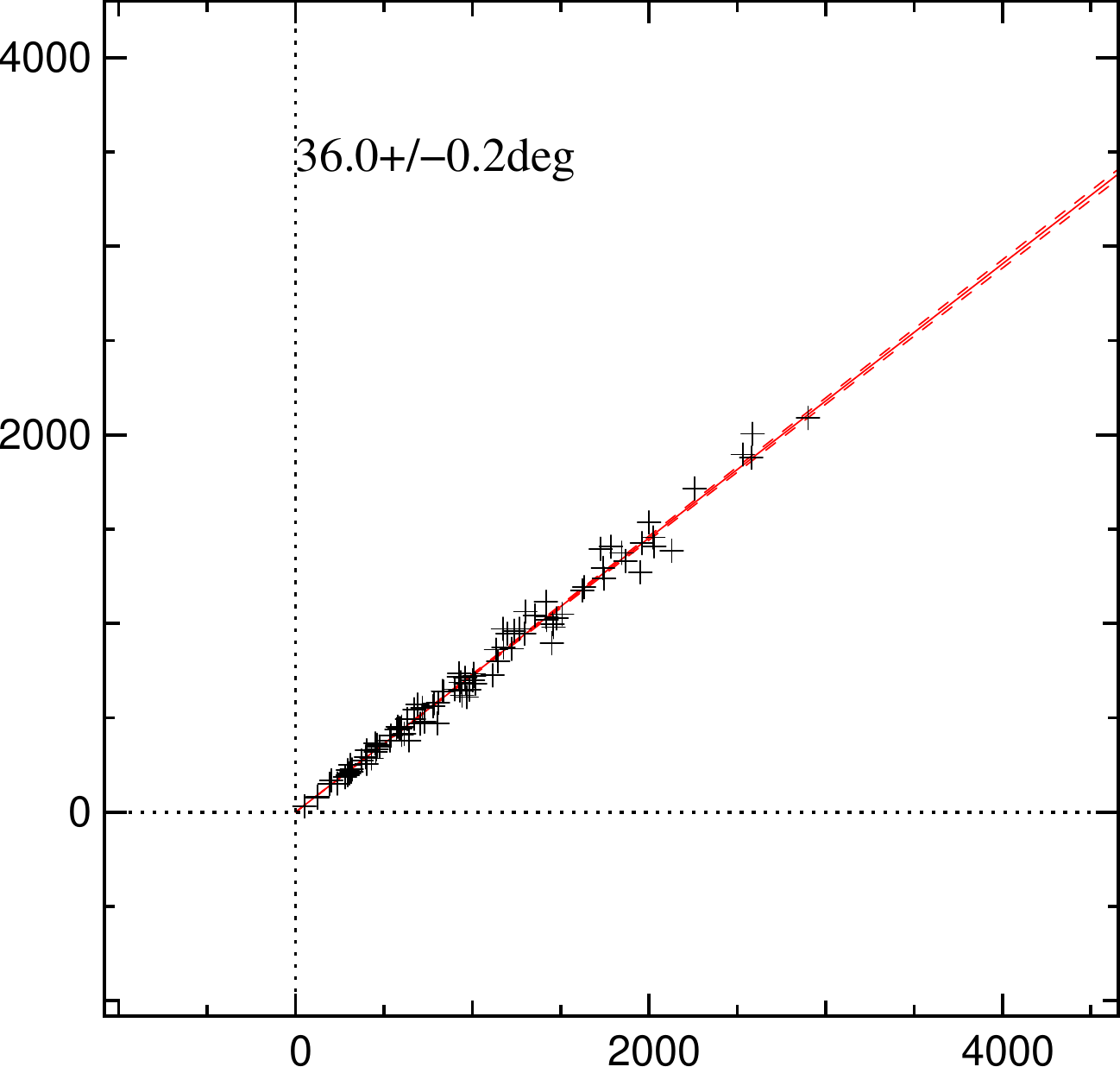}
  \includegraphics[scale=0.345]{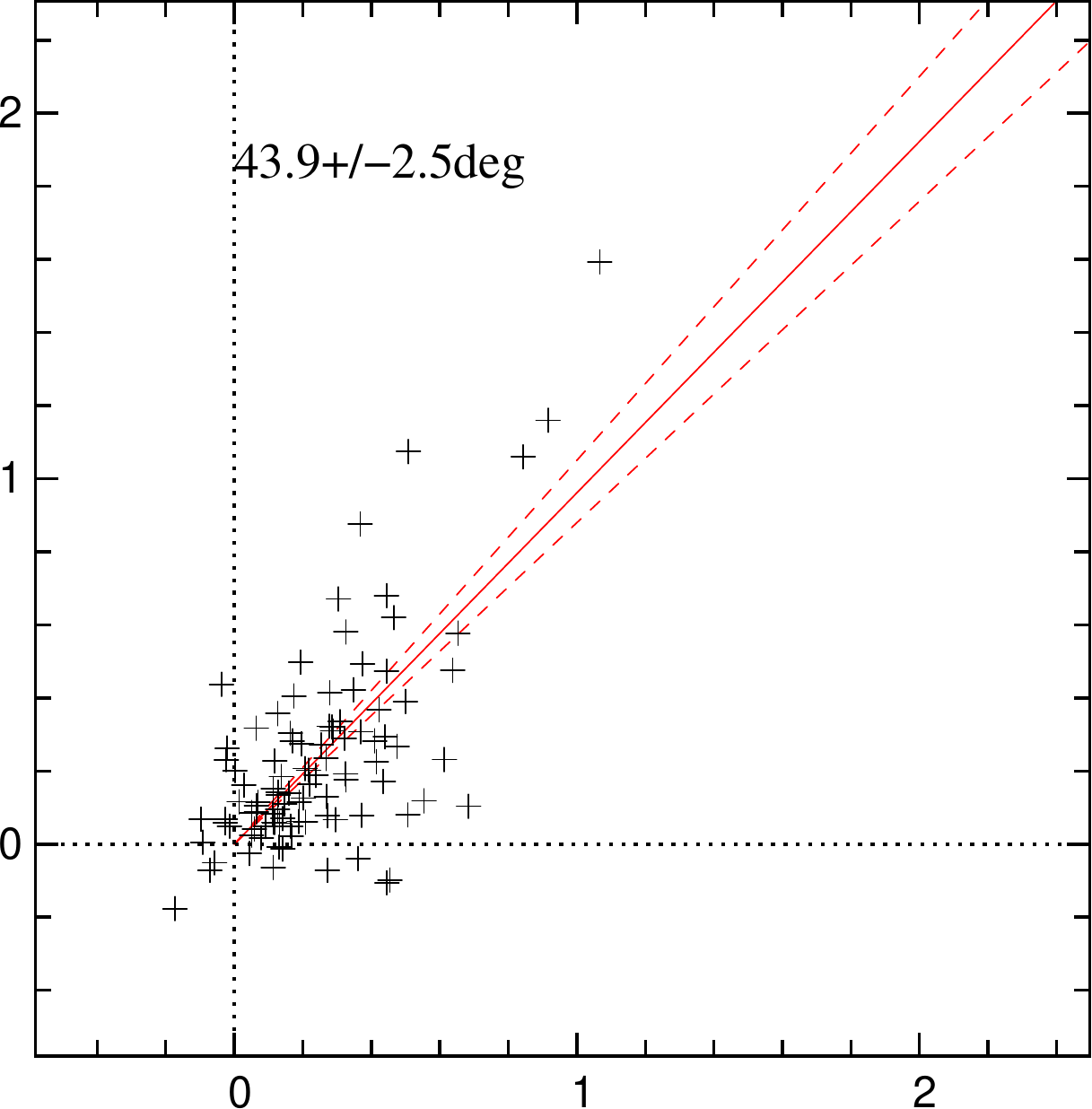}
  \caption{Examples of statistic of the bispectrum over 100 scans obtained in the high SNR (left) and low SNR (right) regimes. The horizontal and vertical axes are the real and imagery parts of the bispectrum respectively (arbitrary unit). The phase of the bispectrum is free from fringe modulation and atmospheric piston terms, and points toward the closure phase. The closure phase average (solid line) and standard deviation (dashed lines) are estimated with the bootstrapping technique. Figure~\ref{fig:interferograms} contains typical interferograms corresponding to these measurements.}
  \label{fig:closure}
\end{figure}

The interferometric signal $m^{ij}$ is filtered in Fourier space in order to keep only the frequencies near the modulation frequency of each baseline. Then the baselines are combined together into bispectra, that are integrated over the length of the scan:
\begin{eqnarray}
b^{ijk} &=& \sum_{opd} m^{ij} \, m^{jk}\, (m^{ik})^*
\end{eqnarray}
Four bispectra are formed $b^{123}$, $b^{124}$, $b^{234}$ and $b^{134}$. Figure~\ref{fig:closure} shows examples of bispectrum statistics obtained for 100 scans in the high SNR and low SNR cases. The phase terms related to the fringe modulation and the atmospheric piston cancel out and all bispectrum estimates point toward the same phase: the closure phase $\Phi^{ij}+\Phi^{jk}-\Phi^{ik}$. We take the argument of the bispectrum average to estimate the final closure phase. We use a bootstraping method to compute its standard deviation.

\begin{figure*}
  \centering
  \includegraphics[width=0.9\textwidth]{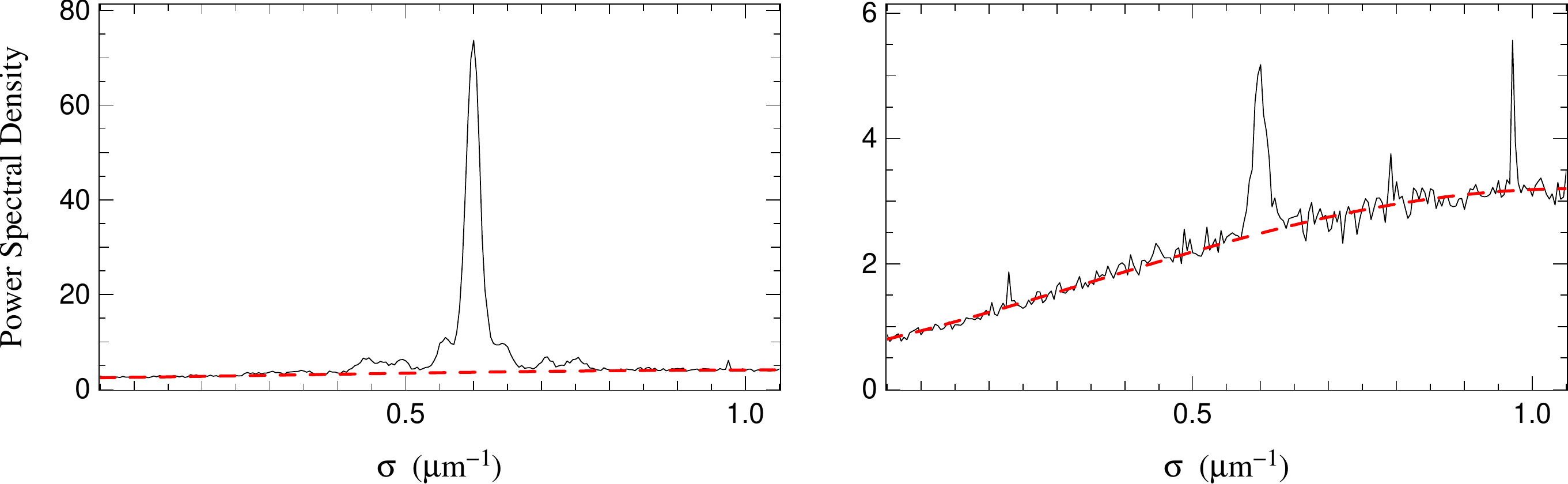}
  \caption{Examples of average Power Spectral Density of the interferometric signals (100 scans) obtained in the high SNR (left) and low SNR (right) regimes. Only a single spectral channel of one baseline is represented. The fringe signal corresponds to the peak at $\sigma=0.6\,\mu$m$^{-1}$ ($\lambda=1.66\,\mu$m). The dashed red line is the fit of the underlying bias power by its theoretical form $a+b\sin^2(2\pi\sigma/\sigma_{max})$.  In the low SNR regime, electronic artefacts can be seen at $\sigma\approx0.22\,\mu$m$^{-1}$,  $0.8\,\mu$m$^{-1}$ and $0.97\,\mu$m$^{-1}$. In the high SNR regime, the fringe peak is clearly aliased by mechanical vibrations in VLTI (see Sec.~\ref{sec:performances}). Figure~\ref{fig:interferograms} contains typical interferogram corresponding to these measurements.}
  \label{fig:psd}
\vspace{0.5cm}
  \centering
  \includegraphics[width=0.9\textwidth]{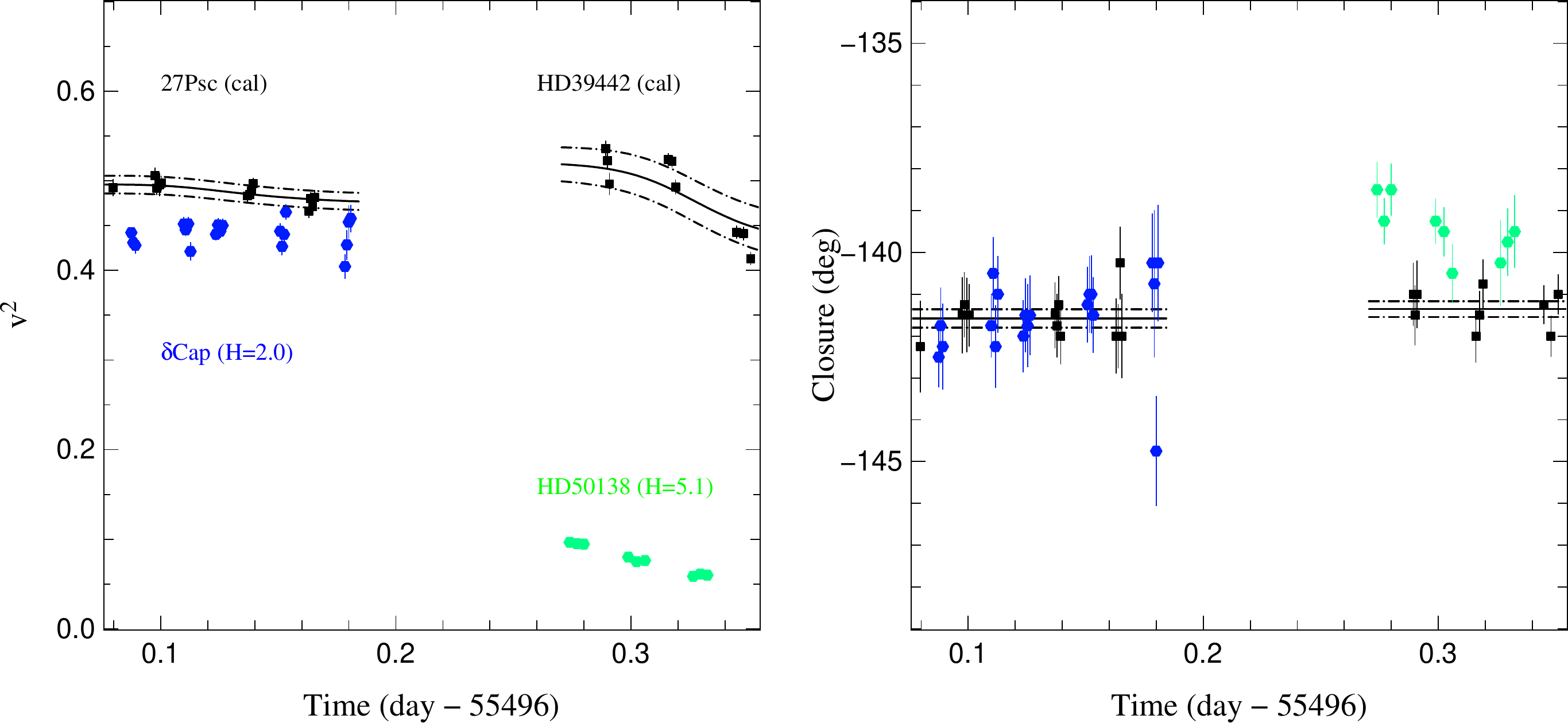}
  \caption{Examples of transfer function calibration for the visibility (left) and the closure phase (right).  Only a single spectral channel and a single baseline (triplet for the closure phase) is represented. The black squares are the transfer function estimates made on calibration stars extracted from the \citet{merand:2005apr} catalogue. The solid and dashed lines represent the interpolation of the transfer function, with its uncertainty. The colored circles are the observations of scientific targets.}
  \label{fig:tf}
\end{figure*}

\subsection{Visibility estimate}
The critical step is the photometric normalization of the interferometric signal $m^{ij}$ by the photometric signal $\sqrt{P^iP^j}$. We implement two versions.

For bright stars, the photometric normalization is performed independently for each step in scan. This simultaneity is the key to obtain precisely calibrated interferograms on bright stars. As discussed by \citet{coude-du-foresto:1997feb} and \citet{kervella:2004oct:b}, we use a Wiener filtering of $\sqrt{P^iP^j}$. Still this is not enough to avoid numerical instabilities when $\sqrt{P^iP^j}$ get close to zero. We filter out the scans with such a behavior.

For fainter stars, the photometric signal is averaged over the length of a scan. When dominated by the detector noise, it dramatically enhances the precision on the photometric signal. In a sense this is an extreme case of the previous method, when the Wiener filtering is so strong that it is equivalent to a simple time average.

After the photometric normalization, we use the method detailled by \citet{coude-du-foresto:1997feb}, applied independently to each spectral channel and baseline, to obtain an estimate of the square visibility $|V^{ij}|^2$. This method is based on the Power Spectral Density (PSD) of the signal to disentangle the fringe power (peak) from the detector and photon noise biases (continuum). Typical PSD are presented in Fig.~\ref{fig:psd}. The shape of the underlying bias power can be correctly fitted with the form  $a+b\sin^2(2\pi\sigma/\sigma_{max})$ where $\sigma{}$ is the wavenumber. The first term $a$ corresponds to the photon noise and the second term $b$ corresponds to the detector noise. This underlying structure of the noise is due to the ``fowler'' readout scheme: the detector noise is high-pass filtered by the difference between two consecutive reads.

\subsection{Transfer function calibration}
The instrumental and atmospheric contributions to the visibilities and closure phases (the so-called transfer function) are monitored by interleaving the observation of science targets with calibration stars. We implemented several strategies to interpolate the transfer function at the observing time of the science targets. We found that the most robust is to simply average all the transfer function estimates of the night, with a weight inversely related to the time separation. The uncertainty on the transfer function at the time of the scientific observations takes into account both the error bars and the dispersion of the transfer function estimates: a sequence of precise (small error bars) but dispersed points leads to large uncertainties on the calibrated product. An example of a transfer function with its interpolation is displayed in Fig.~\ref{fig:tf}.

Calibration stars for bright targets are generally extracted from the catalogues of \citet{merand:2005apr} and \citet{borde:2002oct}. Calibration stars for fainter targets are extracted from the JSDC catalog from JMMC \citep{Lafrasse:2010uq}.

\subsection{Performances}
\label{sec:performances}
In this section, the term \emph{median atmospheric conditions} refers to seeing$\,\approx1''$, $\tau_0\approx3\,$ms, and wind speed$\,\approx6\,$m/s. 

The limiting magnitude of PIONIER is defined by its capability to detect the fringes during the observations (SNR$\approx$$3$ per scan). The typical measured sensitivity is $H\mathrm{mag}\approx5$ for a dispersion over 7 spectral channels and using a scan of 512 steps with the detector in mode ``fowler''. When the atmospheric conditions are better than the median, it is possible to reach a limiting magnitude of $H\mathrm{mag}\approx7$ using a scan of 1024 steps. In broad band, we have been able to detect and track the fringes on several unresolved stars with $H\mathrm{mag}>8$ under good conditions. These numbers are for the Auxiliary Telescopes.

The statistical uncertainties on the closure phases (error bars in Fig.~\ref{fig:tf}, right) are compatible with the data dispersion. The final accuracy of the calibrated closure phases generally ranges from $5\,$deg to $0.5\,$deg. These uncertainties not only depend on the target brightness but also on the turbulence strength. Sequences with stable closure phases, down to $0.1\,$deg, have been recorded on bright stars under atmospheric conditions better than median. At such a level, new possible biases should be studied such as the dependence of the instrumental closure phase on the spectral type. This has not been done yet.

The measured accuracy on calibrated visibilities ranges from $15$ to $3\%$ depending on the atmospheric conditions. On bright targets, the dispersion of the uncalibrated data is sometimes larger than the statistical uncertainties computed by \texttt{pndrs} (Fig.~\ref{fig:tf}, left). It means that we are sometimes facing some non-stationary biases. Our prime suspects are the mechanical OPD perturbations that could be related to the wind strength, perhaps by shaking the telescope structure. Indeed, when the wind speed is higher than $6\,$m/s, the PSDs recorded by PIONIER are clearly affected by vibrations, as seen in Fig.~\ref{fig:psd} left. These perturbations are typically in the range $10-100\,$Hz, with a typical life-time of a few seconds. This should be further investigated before drawing definite conclusions.

The VINCI and IONIC-3 instruments achieved a visibility accuracy of $1\%$, using a similar instrumental concept to PIONIER \citep[see for instance][]{kervella:2004oct:a}. We plan to dedicate a few observing nights to define the best observing setup to reach the same level of accuracy.

\section{Image reconstruction of \deltaSco{} and HIP11231}
\label{sec:illustration}

In this section we illustrate the imaging capability of PIONIER on the stars \deltaSco{} and HIP11231. \deltaSco{} is a well known bright Be star ($H\mathrm{mag}=2.1$) accompanied by a faint companion on a highly eccentric orbit. HIP11231 is a G1V spectroscopic binary with apparent magnitude $H\mathrm{mag}=4.8$. The PIONIER observations are presented in Appendix~\ref{app:datadeltaSco} and~\ref{app:data} together with their $(u,v)$ plane coverage. \deltaSco{} has been observed once with the intermediate configuration (D0-H0-G1-I1). HIP11231 has been observed at 7 epochs, and with the three AT quadruplets (D0-H0-G1-I1, E0-G0-H0-I1, A0-K0-G1-I1). Data have been processed with the \texttt{pndrs} package.

\subsection{Image reconstruction and model fitting of \deltaSco{}}
We perform model-independent image reconstruction with the MIRA software from \citet{Thiebaut:2008}. The starting point is a delta function at $(0,0)$. The pixel scale is $0.2$mas and the field-of-view is $200\times200$ pixels. In practice, the MIRA algorithm minimizes a joint criterion which is the sum of (1) a likelihood term which measures the compatibility with the data, and (2) a regularization term which imposes priors on the image. The relative weight between these two terms is controlled by a multiplicative factor called  ``hyperparameter''. We use the ``total variation'' regularization associated with positivity constraint as recommended by \citet{Renard:2011}. We set the hyperparameter with a small value of 100, so that the weight of the regularization term is kept small with respect to the fit to the data. It brings some superresolution, at the cost of an increased level for the noise in the image. \corrections{We use the information of all the spectral channels to improve the $(u,v)$ plane coverage (``gray image'' hypothesis).}

\begin{figure}
  \centering
  \includegraphics[scale=0.55]{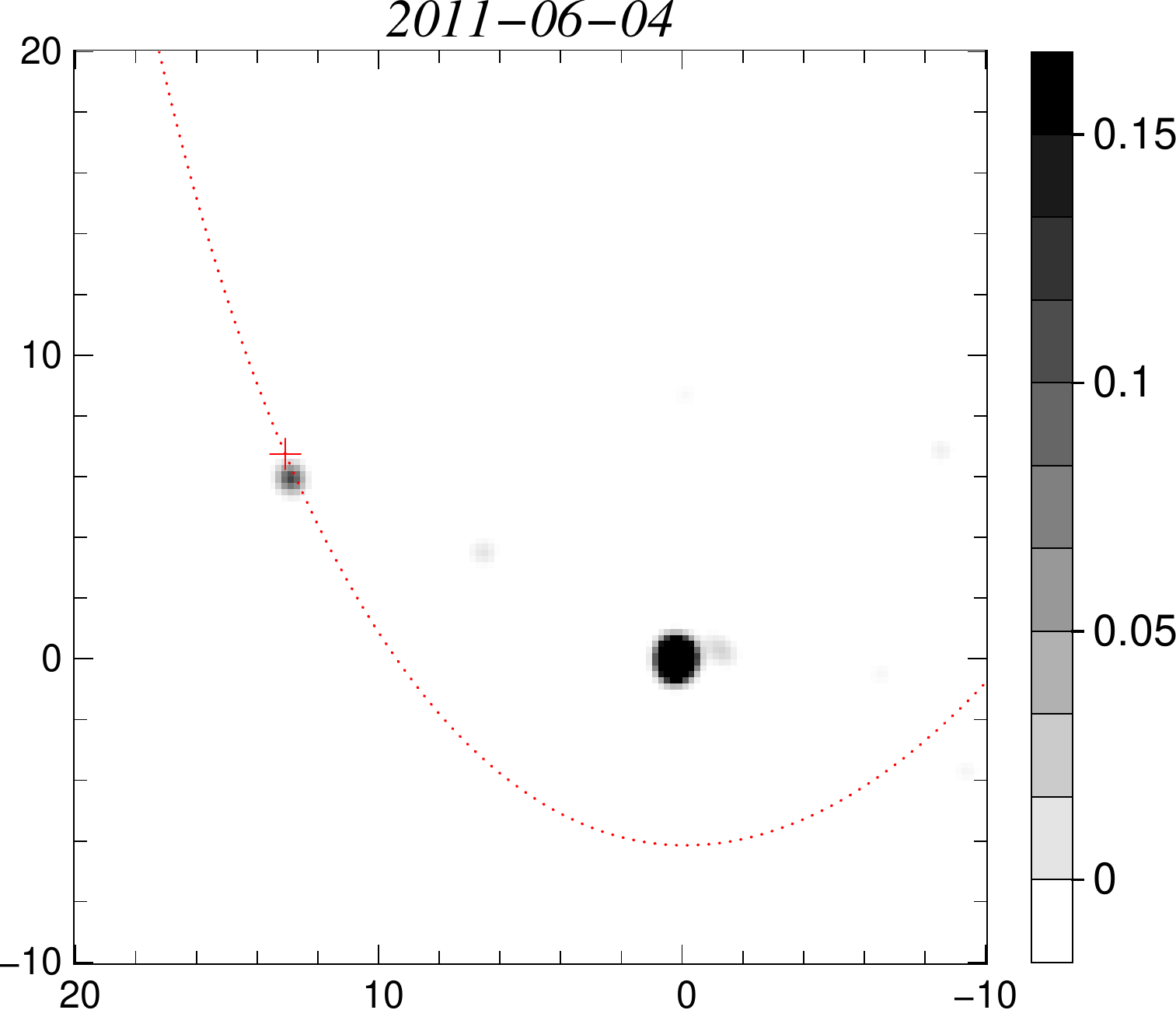}
  \caption{Image reconstruction of \deltaSco{}. The reference frame is centered on the primary star. The red cross is the position of the companion at the time of the PIONIER observations predicted from \citet{Tycner:2011}. Scale is in mas, North is up and East is to the left. The image maximum is normalized to unity and the color scale is linear from $0$ to $0.15$.}
  \label{fig:deltaSco}
\end{figure}

The reconstructed image is presented in Fig.~\ref{fig:deltaSco}. The companion is easily detected. The main component appears extended as the disk surrounding it is marginally resolved by the VLTI baselines. This result illustrates the snapshot imaging capability of PIONIER thanks to the simultaneous beam combination of 4 telescopes.

In parallel, we perform a fit with a binary model with one resolved component. Results are presented in Table~\ref{tab:binaireDeltaSco}. The resulting visibility and closure phase curves are overlaid to the data in Appendix~\ref{app:datadeltaSco}. We checked that the position of the companion is consistent with the prediction by \citet{Tycner:2011}. This image reconstruction of \deltaSco{} demonstrates that the \texttt{pndrs} package computes the sign of the closure phases and of the $(u,v)$ plane in a consistent manner. In other words, it means that the produced OIFITS files are compliant with the sky orientation defined by \citet{Pauls:2005fk}.
\begin{table}
  \centering
  \caption{Best fit disk+point model for \deltaSco{}: Uniform Disk (UD) diameter of the primary, flux ratio ($\rho$) and position ($\delta\mathrm{ra}$ and $\delta\mathrm{dec}$) of the secondary.}
  \begin{tabular}[c]{lcccccccc} \hline \hline \vspace{-0.3cm}\\
    UD & $\rho$ & $\delta\mathrm{ra}$ & $\delta\mathrm{dec}$ & $\chi^2_r$  \\
    $[\mathrm{mas}]$ & $[\%]$ & $[\mathrm{mas}]$& $[\mathrm{mas}]$ &  \vspace{1mm}\\\hline
    \vspace{-3mm}\\
    $1.48\pm0.07$ & $6.3\pm0.5$ & $12.72\pm0.1$ & $5.96\pm0.1$ & 1.1\\\hline
  \end{tabular}
  \flushleft
  \label{tab:binaireDeltaSco}
\end{table}

\begin{figure*}
  \centering
  \includegraphics[width=0.9\textwidth]{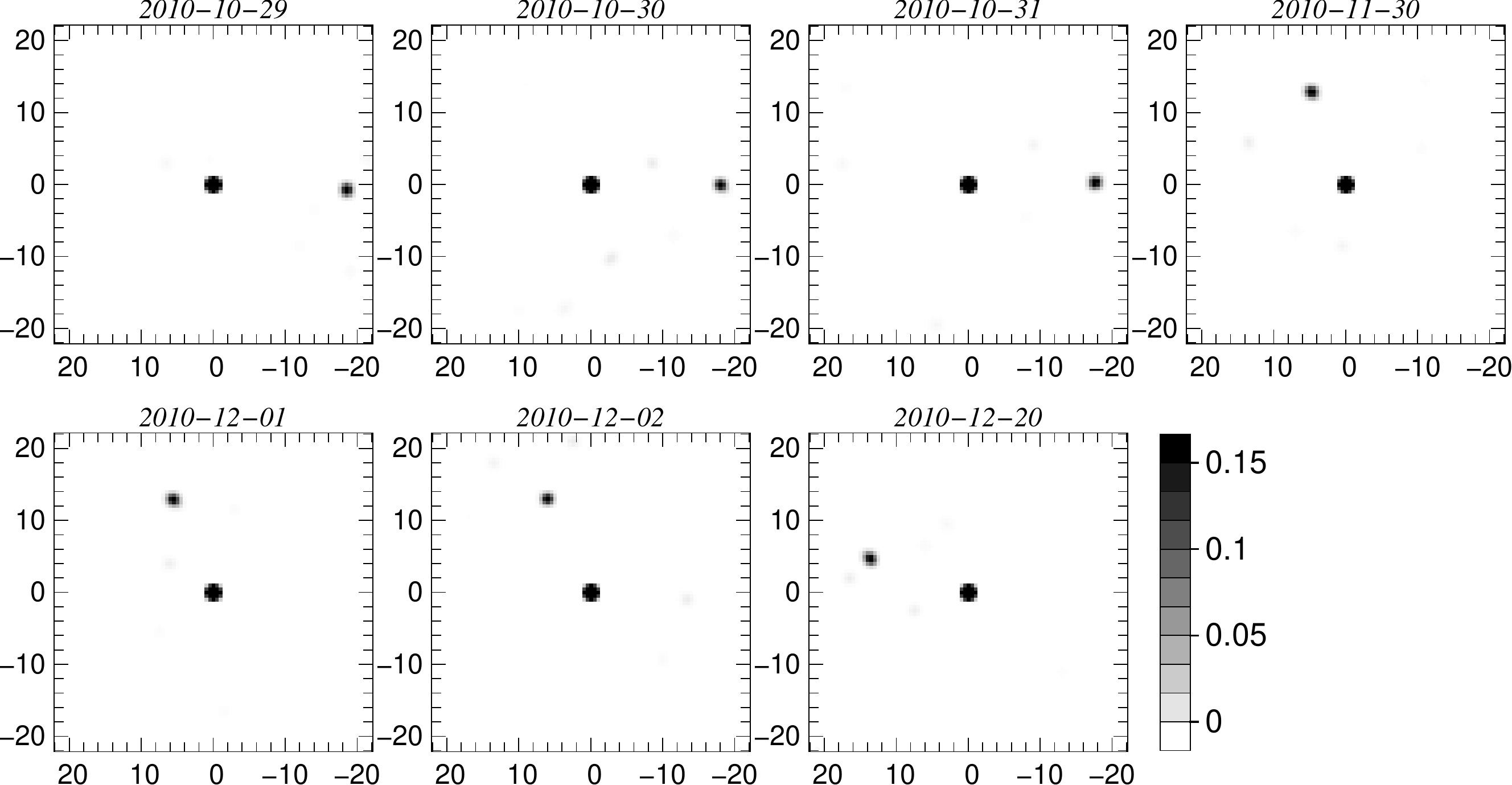}
  \caption{Model independent image reconstruction of HIP11231 with the MIRA software. Each night has been reconstructed independently. The reference frame is centered on the primary star, which saturates the color table. Scale is in mas, North is up and East is to the left. The image maximum is normalized to unity and the color scale is linear from $0$ to $0.15$.}  \label{fig:images}
  \vspace{0.5cm}
  \includegraphics[width=0.85\textwidth]{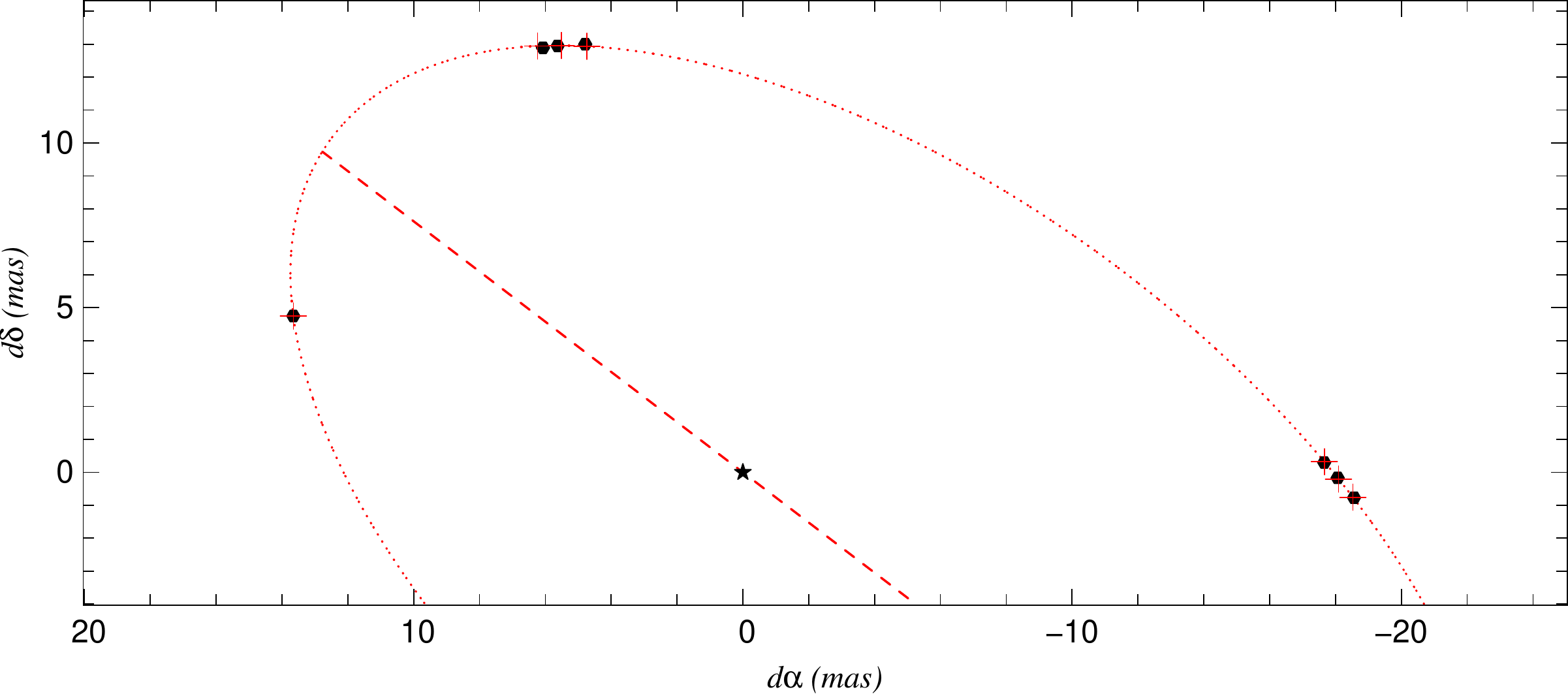}
  \caption{Best fit of the apparent orbit of HIP11231, considering all the Campbell elements free except the period \citep[142.33days, from][]{Barker:1967}. The black circles are the PIONIER astrometric measurements. The red dotted line is the best orbit with its line of nodes in dashed.  The red crosses are the positions of the companion at the time of the PIONIER observations predicted by the best-fit orbit. The average residual distance is $60\,\mu$as. The reference frame is centered on the primary star. North is up and East is to the left.}
  \label{fig:orbit}
\end{figure*}

\subsection{Image reconstruction, model fitting and orbital solution of HIP11231}
We use the same MIRA parameters as for \deltaSco{} except that the pixel scale is set to $0.5$mas. The reconstruction is performed independently for each epoch. \corrections{We use the information of all the spectral channels to improve the $(u,v)$ plane coverage (``gray image'').} Reconstructed images are presented in Fig.~\ref{fig:images}. The spectroscopic binary is easily resolved by PIONIER+VLTI with the three AT configurations. The noise in the images is kept below 2.5\% of the image maximum. The flux ratio between the primary and the secondary, measured by aperture photometry in each reconstructed image, ranges from $\rho=0.18$ to $\rho=0.21$.

In parallel, we perform a global fit to the full data set with the following parameters: (1) the flux ratio, forced to be identical for all epochs and (2) the relative positions, that we let free between each epoch. We combined together all the spectral channels to improve the $(u,v)$ plane coverage. Such a global fit is conveniently performed with the LITpro software \citep{Tallon-Bosc:2008}. The best-fit flux ratio is $\rho=0.206$. At each epoch, the relative separation could be retrieved unambiguously. The fitted positions and the flux ratio are consistent with the results from model-independent image reconstruction. The resulting visibility and closure phase curves are overlaid to the data in Appendix~\ref{app:data}.
\begin{table}[t]
  \centering
  \caption{Campbell orbital elements of HIP11231 from our PIONIER astrometric measurements and from \citet{Jancart:2005}. Both solutions adopt the spectroscopic period $P=142.33$d \citep{Barker:1967}. }
  \begin{tabular}[c]{lcccccccc} \hline \hline \vspace{-0.3cm}\\
    Parameters & $T_0$ & $a^+$ & $e$ & $i$ & $\Omega$ & $\omega$ \\
    & $[\mathrm{d}]$ & $[\mathrm{mas}]$ & & $[^\circ]$ &  $[^\circ]$ & $[^\circ]$\vspace{1mm}\\\hline
    \vspace{-3mm}\\
    Jancart, 2005: & 37159.1 & 8.09 & 0.29 & 60.6	& 200.2 & 188.2 \\
    This work:     & 37187.4 & 21.3 & 0.25 & 57.5 & 232.6 & 194.8 \\\hline
  \end{tabular}
  \flushleft
  {\footnotesize $^+$ Semi-major axis of the barycenter of light in \citet{Jancart:2005}, semi-major axis of secondary orbit in this work.}
  \label{tab:orbit}
\end{table}

Figure~\ref{fig:orbit} shows the relative separations derived from the fit overlaid with the best orbital solution. We adopt the spectroscopic period derived by \citet{Barker:1967} from radial velocity, but we allowed all the other Campbell elements to vary, namely: the time passage through periastron $T_0$, the eccentricity $e$, the semi-major axis $a$, the position angle of the ascending node $\Omega$, the argument of periastron $\omega$ and the orbital inclination $i$. Parameters of the best orbital solution are presented in Table~\ref{tab:orbit} together with the solution determined from Hipparcos astrometry by \citet{Jancart:2005}. The difference in semi-major axis $a$ is due to the fact that \citet{Jancart:2005} reconstructed the orbit of the barycenter of light from unresolved astrometry, while we reconstruct the orbit of the secondary from resolved astrometry. The difference in time passage through periastron $T_0$ can be explained by the uncertainties on the period and by the fact that our observations have been taken several orbital periods later than the Hipparcos ones. The disagreement in $\Omega$ is more striking as we cannot find a simple explanation.

\section{Conclusions and perspectives}

\corrections{The 4-telescope PIONIER instrument has been successfully developed and commissioned at the VLTI interferometer. The use of integrated optics technologies combined with the expertise gained from projects such as FLUOR, AMBER, VINCI and IONIC-3 made it possible to design, build and commission the instrument in about 18 months. PIONIER is routinely used for science operations in the H-band and permits high angular resolution imaging studies at an unprecedented level of sensitivity and precision. The overall performances are in agreement with the requirements from the key programs, although some work is still needed to achieve the best possible accuracy on the calibrated visibilities and closure phases.}

PIONIER is expected to stay a few years at VLTI as a visitor instrument, until the arrival of the second generation instruments MATISSE and GRAVITY. Within this lifetime, several improvements are contemplated:
\begin{itemize}
\item The imaging and dispersion unit will be upgraded with a motorized translation stage for the dispersion prisms. It will allow to change the operational mode (broad band or dispersed) during the night. This should be installed in PIONIER around December 2011.
\item Changing the operational wavelength from the H-band to the K-band would benefit all the key science programs. A component is already available at IPAG. The design of the associated imaging optics is under study. Ideally, the changes will be installed in PIONIER early 2012.
\item The arrival of a new generation of infrared detectors should provide detector noise as low as $2.5\,e^-$ at a frame-rate larger than $1\,$kHz in destructive mode \citep{Gach:2009,Finger:2010}. Installed in PIONIER, such a detector would increase the sensitivity by $1.5\,$mag and would solve the readout speed and saturation issues.
\item PIONIER has the status of a visitor-instrument at Paranal Observatory. The possibility to offer PIONIER to the community is currently investigated.
\end{itemize}

\begin{acknowledgements} 
PIONIER is funded by the Universit\'e Joseph Fourier (UJF, Grenoble) through its Poles TUNES and SMING and the vice-president of research, the Institut de Plan\'etologie et d'Astrophysique de Grenoble, the ``Agence Nationale pour la Recherche'' with the program ANR EXOZODI, and the Institut National des Science de l'Univers (INSU) with the programs ``Programme National de Physique Stellaire'' and ``Programme National de Plan\'etologie''. \corrections{The integrated optics beam combiner is the result of a collaboration between IPAG and CEA-LETI based on CNES R\&T funding.} The authors want to warmly thank all the people involved in the VLTI project. This work is based on observations made with the ESO telescopes. It made use of the Smithsonian/NASA Astrophysics Data System (ADS) and of the Centre de Donnees astronomiques de Strasbourg (CDS). All calculations and graphics were performed with the freeware \texttt{Yorick}.
\end{acknowledgements}

~\newpage


\appendix
\begin{table*}[!h]
  \section{Observations of \deltaSco{}}
  \label{app:datadeltaSco}
  \centering
  \caption{Log of the observations of \deltaSco{}. The number of observations is reported taking into account the spectral dispersion (3 spectral channels).  Each measurement (pointing) is generally the average of 3 or 5 files of 100 scans each.}
  \begin{tabular}[c]{lccrrrccc} \hline \hline \vspace{-0.3cm}\\
    Date &Average MJD & Baseline &\# of $V^2$ & \# closure &  Time spend  \\\hline
    2011-06-04 & 55717.2 & D0-H0-G1-I1 & 24$\times$3  & 16$\times$3 & 1.8h  \\\hline
  \end{tabular}\vspace{1cm}
\end{table*}
\begin{figure*}[!h]
  \centering
  \includegraphics[width=0.9\textwidth]{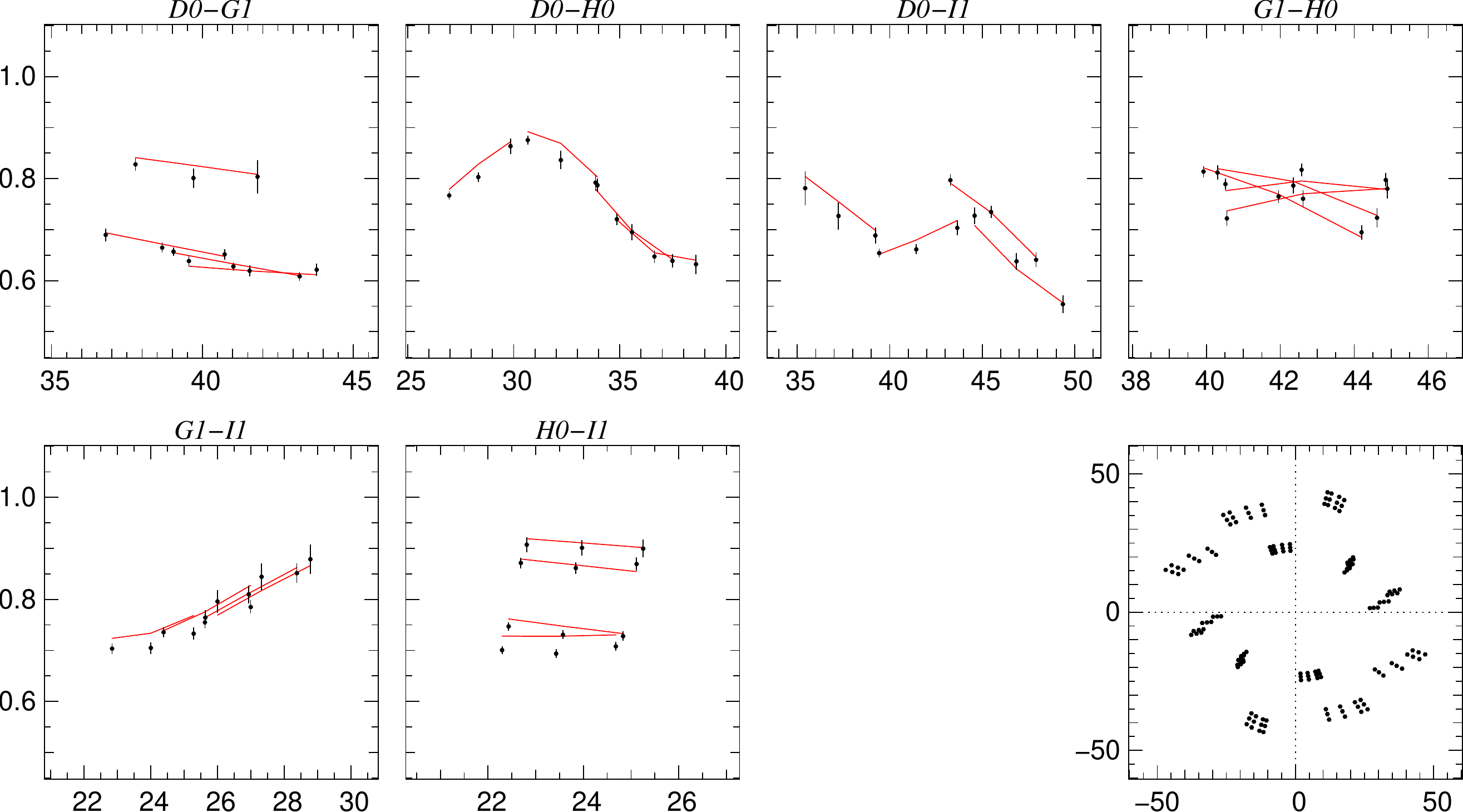}
  \caption{Observed calibrated visibilities ($V^2$) on \deltaSco{} versus the spatial frequency (m$/\mu$m). Each panel is for a baseline. The red curves are the result of the global fit by a disk+point model. The $(u,v)$ plane coverage (m$/\mu$m) of the observations is displayed in the bottom-right panel. North is up and East is to the right. The radial tracks come from the spectral dispersion over 3 spectral channels.}
  \label{fig:v2ObsDeltaSco}
  \vspace{1cm}
\end{figure*}
\begin{figure*}[!h]
  \centering
  \includegraphics[width=0.9\textwidth]{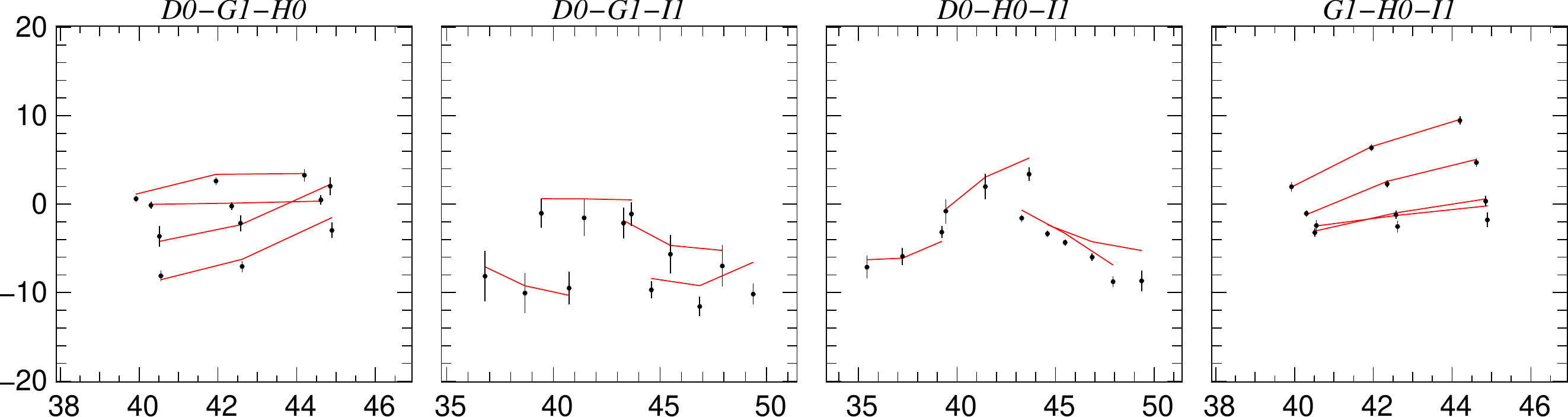}
  \caption{Observed calibrated closure phase (deg) on \deltaSco{} versus the largest spatial frequency of the baseline triple (m$/\mu$m). Each panel is for a baseline triplet. The red curves are the result of the global fit by a disk+point model.}
  \label{fig:t3phiObsDeltaSco}
\end{figure*}

\clearpage
\begin{table*}[!t]
  \section{Observations of HIP11231}
  \label{app:data}
  \centering
  \caption{Log of the observations of HIP11231. The number of observations is reported taking into account the spectral dispersion (7 spectral channels). Each measurement (pointing) is generally the average of 3 or 5 files of 100 scans each.}
  \begin{tabular}[c]{lccrrrccc} \hline \hline \vspace{-0.3cm} \\
    Date &Average MJD & Baseline &\# of $V^2$ & \# closure &  Time spend  \\\hline
    2010-10-29 & 55499.2 & D0-H0-G1-I1 & 24$\times$7  & 16$\times$7 & 2.2h  \\
    2010-10-30 & 55500.2 & D0-H0-G1-I1 & 18$\times$7  & 12$\times$7 & 1.4h  \\
    2010-10-31 & 55501.1 & D0-H0-G1-I1 & 18$\times$7  & 12$\times$7 & 1.2h  \\
    2010-11-30 & 55531.1 & E0-G0-H0-I1 & 12$\times$7 & 8$\times$7 & 1h \\
    2010-12-01 & 55532.1 & E0-G0-H0-I1 & 12$\times$7  & 8$\times$7 & 0.8h \\
    2010-12-02 & 55533.0 & E0-G0-H0-I1 & 12$\times$7 & 8$\times$7 & 0.9h \\
    2010-12-20 & 55551.1 & A0-K0-G1-I1 & 12$\times$7 & 8$\times$7 & 0.65h \\\hline
  \end{tabular}\vspace{1cm}
\end{table*}
\begin{figure*}[!h]
  \centering
  \includegraphics[width=0.9\textwidth]{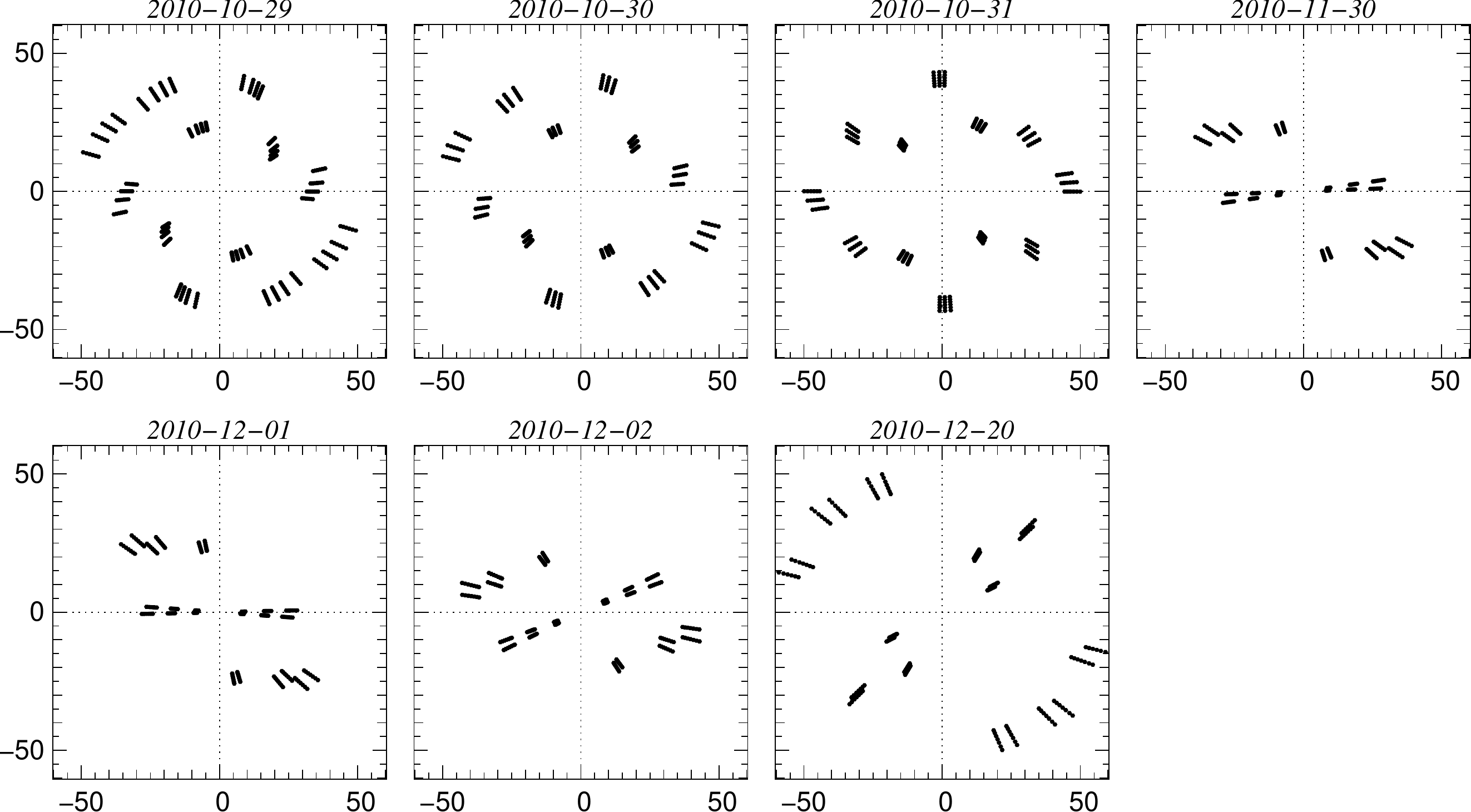}
  \caption{$(u,v)$ plane coverage (m$/\mu$m) of the observations of HIP11231. North is up and East is to the right. The radial tracks come from the spectral dispersion over 7 spectral channels. Each panel is for an epoch.}
  \label{fig:uvObs}
\end{figure*}\vspace{1cm}

\begin{figure*}[!h]
  \centering
  \includegraphics[width=0.9\textwidth]{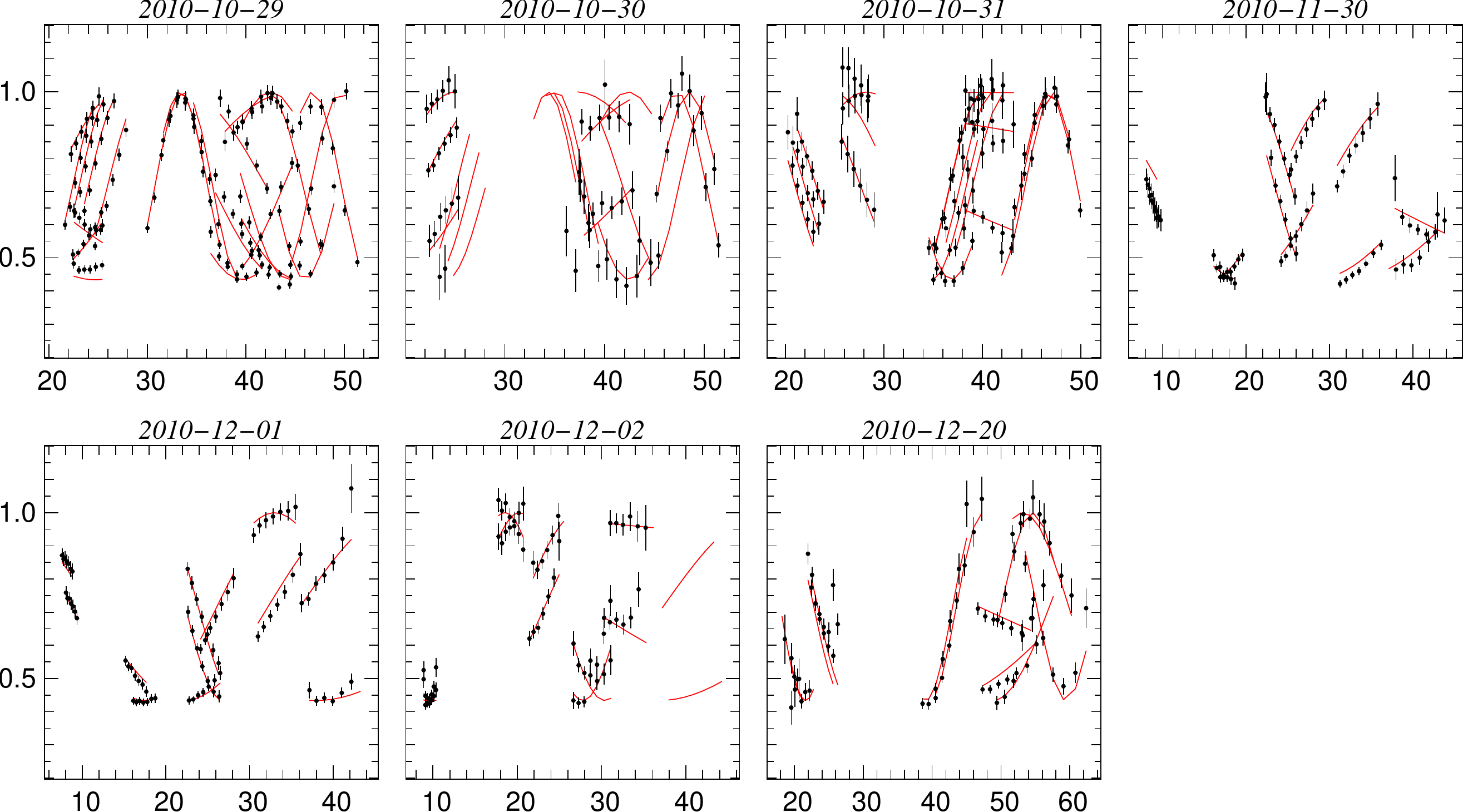}
  \caption{Observed calibrated visibilities ($V^2$) on HIP11231 versus the spatial frequency (m$/\mu$m). The red curves are the result of the global fit by a binary model. Each panel is for an epoch.}
  \label{fig:v2Obs}\vspace{1cm}
\end{figure*}
\begin{figure*}[!h]
  \centering
  \includegraphics[width=0.9\textwidth]{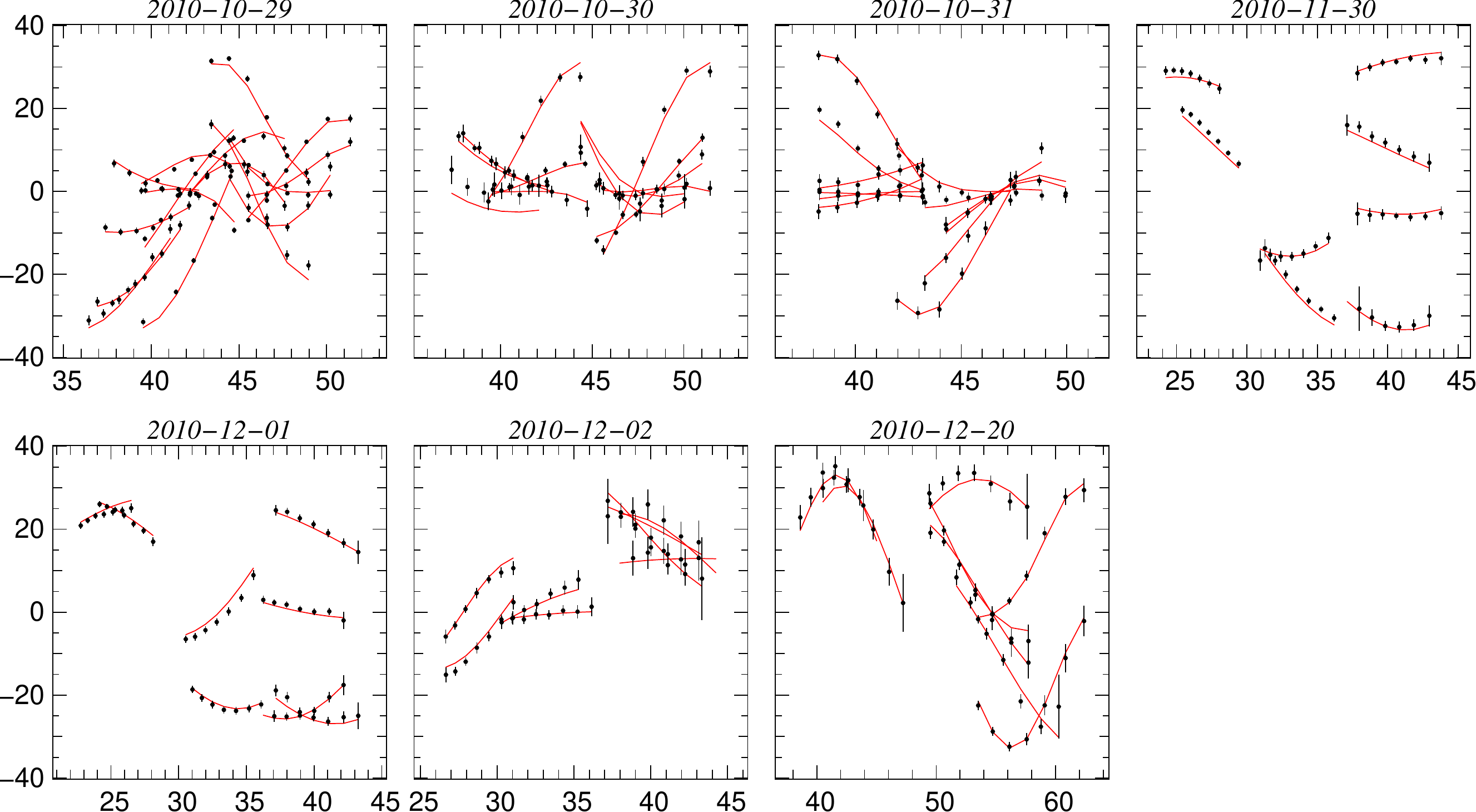}
  \caption{Observed calibrated closure phase (deg) on HIP11231 versus the largest spatial frequency of the baseline triple (m$/\mu$m). The red curves are the result of the global fit by a binary model. Each panel is for an epoch.}
  \label{fig:t3phiObs}
\end{figure*}

\end{document}